\documentclass[aps,prb,twocolumn,showpacs,groupaddress,floatfix]{revtex4}
\usepackage{graphicx,graphics}
\usepackage{dcolumn}
\usepackage{amsmath,amssymb,amsfonts}
\usepackage{latexsym,verbatim}
\usepackage{bm}
\usepackage{color}
\usepackage{ulem}
\usepackage[breaklinks=true,colorlinks,citecolor=blue,linkcolor=blue,urlcolor=blue]{hyperref}
\def\be{\begin{equation}}
\def\ee{\end{equation}}
\begin{document}
\title{Intrinsic lifetime of Dirac plasmons in graphene}
\author{Alessandro Principi}
\email{principia@missouri.edu}
\affiliation{Department of Physics and Astronomy, University of Missouri, Columbia, Missouri 65211, USA}	
\author{Giovanni Vignale}
\affiliation{Department of Physics and Astronomy, University of Missouri, Columbia, Missouri 65211, USA}
\author{Matteo Carrega}
\affiliation{NEST, Istituto Nanoscienze-CNR and Scuola Normale Superiore, I-56126 Pisa, Italy}
\author{Marco Polini}
\affiliation{NEST, Istituto Nanoscienze-CNR and Scuola Normale Superiore, I-56126 Pisa, Italy}

\begin{abstract}
Dirac plasmons in a doped graphene sheet have recently been shown to enable confinement of light to ultrasmall volumes. In this work we calculate the {\it intrinsic} lifetime of a Dirac plasmon in a doped graphene sheet by analyzing the role of electron-electron interactions beyond the random phase approximation. The damping mechanism at work is intrinsic since it operates also in disorder-free samples and in the absence of lattice vibrations. We demonstrate that graphene's sublattice-pseudospin degree of freedom suppresses intrinsic plasmon losses with respect to those that occur in ordinary two-dimensional electron liquids. We relate our findings to a microscopic calculation of the homogeneous dynamical conductivity at energies below the single-particle absorption threshold. 
\end{abstract}
\maketitle

\section{Introduction}
Plasmons are ubiquitous high-frequency collective density oscillations of an electron liquid, which occur both in metals and insulators~\cite{Pines_and_Nozieres,Giuliani_and_Vignale}.  The study of optical phenomena in the nanoscale vicinity of metal surfaces, {\it i.e.} {\it nanoplasmonics}~\cite{Maier,plasmonics}, revolves around the coupling between light and plasmons, which, in turn, enables the compression of electromagnetic energy to the nanometer scale of modern electronic devices.
Of particular interest for novel applications is the study of  the so-called ``Dirac plasmons" (DPs)~\cite{DiracplasmonsRPA,abedinpour_prb_2011,orlita_njp_2012,levitov_arxiv_2013} of the two-dimensional (2D) electron liquid in a doped graphene sheet~\cite{novoselov_naturemater_2007,castroneto_rmp_2009,kotov_rmp_2012} where the carriers are massless Dirac fermions (MDFs).  The properties of DPs have been studied experimentally by a variety of spectroscopic methods~\cite{grigorenko_naturephoton_2012} and  their coupling to infrared light has been engineered in a number of ways~\cite{ju_naturenano_2011,fei_nanolett_2011,fei_nature_2012, chen_nature_2012,yan_naturenano_2012,vicarelli_naturemater_2012}. 
These experiments have revealed that the plasmon wavelength can be much smaller than the illumination wavelength and that DP properties are easily gate-tunable,  thus igniting the field of  ``graphene plasmonics"~\cite{grigorenko_naturephoton_2012}.

Mathematically, a plasmon is an isolated pole $\Omega_{\rm p}(q) = \omega_{\rm p}(q) - i \Gamma_{\rm p}(q)$ of the density-density linear-response function, $\chi_{nn}(q,\omega)$,~\cite{Pines_and_Nozieres,Giuliani_and_Vignale} located slightly below the real axis, $0< \Gamma_{\rm p}(q) \ll \omega_{\rm p}(q)$. The real part of the DP dispersion relation $\omega_{\rm p}(q)$ displays the usual dependence $\omega_{\rm p}(q) \propto \sqrt{q}$ on wave vector $q$~\cite{DiracplasmonsRPA}, typical of 2D electron liquids~\cite{Giuliani_and_Vignale}. The prefactor, however, displays certain peculiarities stemming from broken Galilean invariance~\cite{abedinpour_prb_2011}. By using perturbation theory to first order in electron-electron (e-e)  interactions~\cite{abedinpour_prb_2011}, it has been shown that the prefactor of the plasmon dispersion at long wavelengths is controlled by an interaction-enhanced Drude weight~\cite{abedinpour_prb_2011,orlita_njp_2012}.    The long-wavelength DP dispersion has also been analyzed within Landau theory of Fermi liquids~\cite{levitov_arxiv_2013}. The results of Refs.~[\onlinecite{abedinpour_prb_2011,levitov_arxiv_2013}] cannot be obtained on the basis of the random phase approximation (RPA)~\cite{Pines_and_Nozieres,Giuliani_and_Vignale}.

A key figure of merit of nanoplasmonics is the plasmon lifetime $\tau_{\rm p}(q) = [2 \Gamma_{\rm p}(q)]^{-1}$, or, equivalently, the inverse quality factor $\gamma_{\rm p}(q) = \Gamma_{\rm p}(q)/\omega_{\rm p}(q)$. Plasmon damping is controlled by e-e, electron-impurity, and electron-phonon scattering. The {\it relative} importance of these mechanisms on the propagation of DPs has not yet been quantified theoretically. Experimentally, Fei~{\it et al.}~\cite{fei_nature_2012} have reported a careful experimental analysis of the DP damping rate, which is found to be substantially larger than that predicted on the basis of the Drude transport time and linked to the large background of absorption~\cite{li_natphys_2008} below the single-particle threshold $\hbar \omega < 2 \varepsilon_{\rm F}$, with $\varepsilon_{\rm F}$ the Fermi energy. More recently, Yan {\it et al.}~\cite{yan_naturephoton_2013} have shown that the damping rate of mid-infrared DPs  is strongly affected by substrate and intrinsic phonons. 

As a first step towards a complete elucidation of the  mechanisms that contribute to the DP  lifetime, in this work we present a theory of the {\it intrinsic} DP lifetime.  By ``intrinsic"  we mean  the contribution to $\tau_{\rm p}$ that is solely determined by e-e collisions and therefore survives also in the complete absence of disorder and lattice vibrations.   
\begin{figure}[t]
\begin{center}
\tabcolsep=0cm
\begin{tabular}{c}
\includegraphics[width=1.0\linewidth]{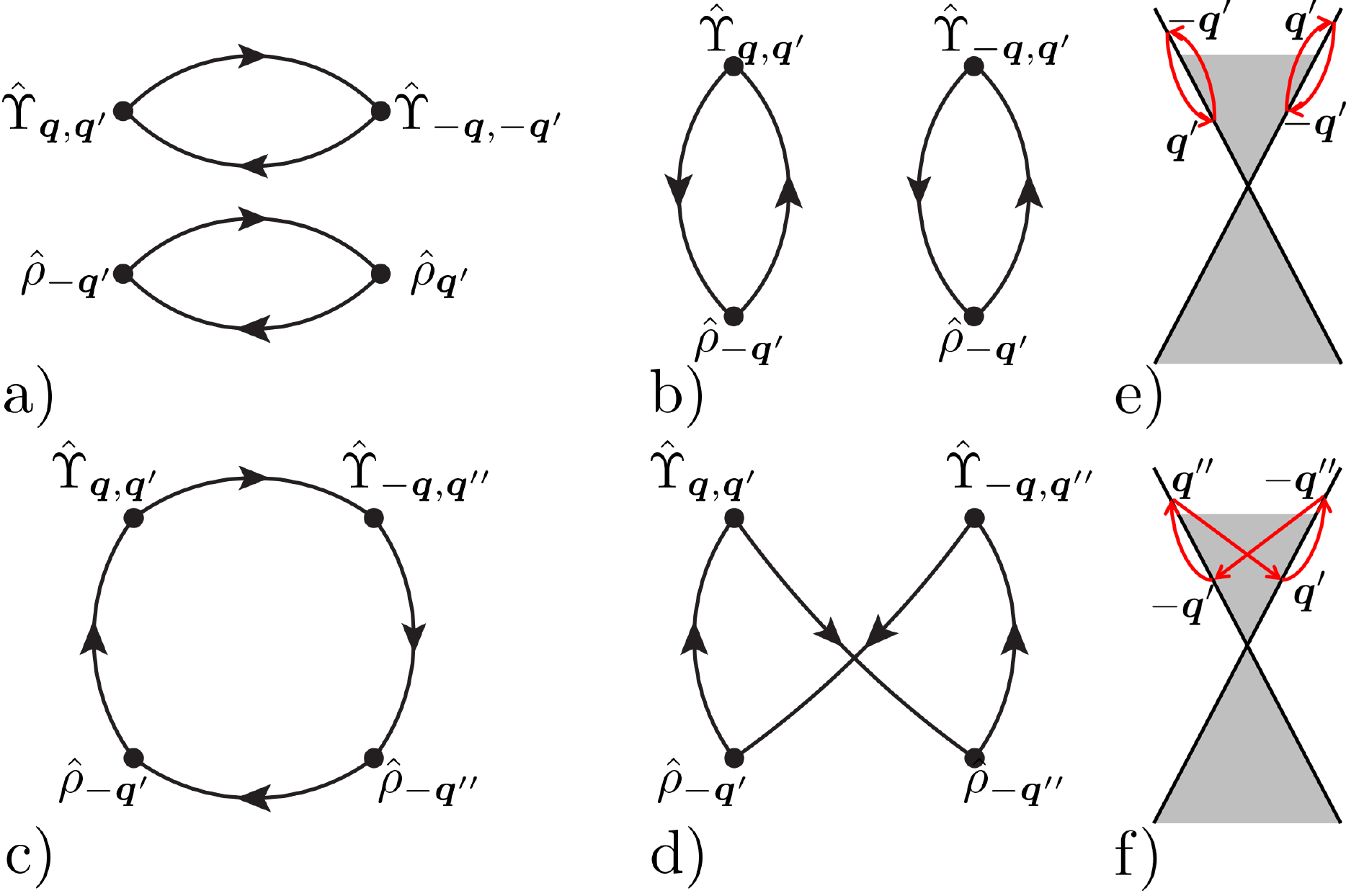}
\end{tabular}
\caption{(Color online) Panels a)-d) show some of the diagrams that contribute to the non-interacting two-particle response function. The diagrams in panels a) and b) are the only two that contribute in the large-N$_{\rm f}$ limit. Panels c) and d) show two non-disconnected diagrams which differ from each other for the order of the external vertices. Panel e)  [panel f)] depicts the excitations that are responsible for the plasmon damping in diagrams a) and b) [c) and d)].\label{fig:one}}
\end{center}
\end{figure}
For extreme concentration of electromagnetic energy the plasmon momenta $q$ of interest are much larger than $q_{\rm light} = \omega_{\rm ph}/c$, where $\hbar\omega_{\rm ph}$ is the free-space photon energy~\cite{fei_nanolett_2011,fei_nature_2012,chen_nature_2012},
but still much smaller than the Fermi wave number $k_{\rm F} = \sqrt{\pi n}$ for typical electron densities~\cite{electronholesymmetry} $n \sim 10^{11}$-$10^{12}~{\rm cm}^{-2}$. 
For $q_{\rm light} \ll q \ll k_{\rm F}$ the DP dispersion satisfies the inequality $\hbar \omega_{\rm p}(q) < 2 \varepsilon_{\rm F}$ and therefore a plasmon {\it cannot} decay by emitting single electron-hole pairs (Fig.~\ref{fig:one}), a mechanism that would be captured by the RPA~\cite{DiracplasmonsRPA}. Therefore, in this regime of momenta, the RPA erroneously predicts no damping whatsoever~\cite{Giuliani_and_Vignale,plasmonlifetimeclassics,NCT}. To correct this, we carry out a calculation of $\tau_{\rm p}$ for DPs in a doped graphene sheet by employing second-order perturbation theory in the strength of e-e interactions. Physically, the lifetime we calculate is determined by decay processes in which a plasmon emits {\it two} electron-hole pairs~\cite{Giuliani_and_Vignale,plasmonlifetimeclassics,NCT}. Our final expression  for $\tau_{\rm p}$ is {\it exact} in the limit of a large number $N_{\rm f}$ of fermion flavors~\cite{Nfgraphene}.   

\section{Theoretical formulation}
The imaginary part $\Gamma_{\rm p}(q)$ of the plasmon dispersion is related to the imaginary part of the density-density response function by the standard formula~\cite{Giuliani_and_Vignale}
\begin{equation}\label{eq:plasmonlifetime} 
\Gamma_{\rm p}(q)= \left.\frac{\Im m[\chi_{nn}(q,\omega)]}{\partial \Re e[\chi_{nn}(q,\omega)]/\partial \omega} \right\vert_{\omega = \omega_{\rm p}(q)}~.
\end{equation}
Since $\Im m[\chi_{nn}(q,\omega)]$ is a quantity of second order in e-e interactions, $\Re e[\chi_{nn}(q,\omega)]$ can be calculated to 
zeroth order in the interaction and the real part of the plasmon frequency $\omega_{\rm p}(q)$  can be taken from the RPA~\cite{DiracplasmonsRPA,grigorenko_naturephoton_2012}: $\omega_{\rm p}(q) = \sqrt{2{\cal D}_0q/\epsilon}$, where ${\cal D}_0 = 4\varepsilon_{\rm F} \sigma_{\rm uni}/\hbar$ is the non-interacting Drude weight, $\sigma_{\rm uni} = N_{\rm f} e^2/(16 \hbar)$ is the so-called universal optical conductivity~\cite{universal}, and $\epsilon = (\epsilon_1 + \epsilon_2)/2$ is the  average of the dielectric constants of the media above ($\epsilon_1$) and below ($\epsilon_2$) the graphene flake~\cite{castroneto_rmp_2009,kotov_rmp_2012}. 

The imaginary part of the density-density response function can now be expressed in terms of the imaginary part of the longitudinal current-current response function $\chi_{jj}(q,\omega)$ according to the equation
\be \label{eq:continuity_equation}
\Im m[\chi_{nn}(q,\omega)] = \frac{q^2}{\omega^2}\Im m[\chi_{jj}(q,\omega)]
\ee
where the longitudinal component of the current density operator $\hat{\bm j}_{\bm q}$ is obtained from the continuity equation for the density operator ${\hat n}_{\bm q}$ (from now on, $\hbar =1$):  $i\partial_t {\hat n}_{\bm q} =[{\hat {\cal H}}, {\hat n}_{\bm q}] =  -{\bm q} \cdot {\hat {\bm j}}_{\bm q}$, with Hamiltonian ${\hat {\cal H}} = {\hat {\cal H}}_0 + {\hat {\cal H}}_{\rm ee}$, where ${\hat {\cal H}}_0$ is the non-interacting Hamiltonian, while ${\hat {\cal H}}_{\rm ee}$ describes Coulomb interactions between density fluctuations.  For ${\hat {\cal H}}_0$ we use the graphene tight-binding (TB) Hamiltonian with nearest-neighbor hopping~\cite{castroneto_rmp_2009} rather than the MDF low-energy effective model~\cite{novoselov_naturemater_2007,castroneto_rmp_2009,kotov_rmp_2012}.  The low-energy MDF limit is taken only {\it after} carrying out all the necessary commutators. In view of this limit, we wrote Eq.~(\ref{eq:continuity_equation}) for a translationally invariant and isotropic system. By following this procedure we avoid problems associated with the ultraviolet cut-off, which breaks gauge invariance~\cite{abedinpour_prb_2011} and is responsible for the appearance of anomalous commutators~\cite{abedinpour_prb_2011,sabio_prb_2008}. See also in appendix.

To proceed, we introduce a unitary transformation generated by a Hermitian operator ${\hat F}$:
$
{\hat {\cal H}}' = e^{i {\hat F}}{\hat {\cal H}}e^{-i {\hat F}}
$,
where the operator ${\hat F}$ is chosen in such a way as to cancel e-e interactions from the transformed Hamiltonian, {\it i.e.} to have ${\hat {\cal H}}' \equiv {\hat {\cal H}}_0$.  This can be done systematically order-by-order in perturbation theory, by expanding ${\hat F} = {\hat \openone} + {\hat F}_1 + {\hat F}_2 +...$, where ${\hat \openone}$ denotes the identity and ${\hat F}_n$ denotes the $n$-th order term in powers of the strength of e-e interactions. We obtain a chain of equations connecting ${\hat F}_n$ to ${\hat {\cal H}}_{\rm ee}$.  For example, to eliminate e-e interactions up to first order, ${\hat F}_1$ must obey the equation $i[{\hat F}_1, {\hat {\cal H}}_0] + {\hat {\cal H}}_{\rm ee} =0$, which can be easily solved~(see App.~\ref{sect:SM_calculation_F1_j1}). 

Note that after carrying out the transformation ${\hat F}$, both the ground state of ${\hat {\cal H}}'$  and the time evolution of the Heisenberg operator becomes non-interacting. 
This is clearly a big simplification. The transformed current operator, 
$
{\hat {\bm j}}'_{\bm q} = e^{i {\hat F}}{\hat {\bm j}}_{\bm q}e^{-i {\hat F}} 
$,
however, becomes complicated. 

The key idea now is to realize that  the calculation of $\Im m[\chi_{j'j'}(q,\omega)]$
 to {\it second} order in the strength of e-e interaction requires only the knowledge of the transformed current-density operator ${\hat {\bm j}}'_{\bm q}$ to {\it first} order, {\it i.e.} ${\hat {\bm j}}'_{\bm q} = {\hat {\bm j}}_{\bm q} + {\hat {\bm j}}_{1, {\bm q}}$, where ${\hat {\bm j}}_{1, {\bm q}}=i[{\hat F}_1, {\hat {\bm j}}_{{\bm q}}] $.
The untransformed current operator ${\hat {\bm j}}_{\bm q}$ is indeed a one-particle operator and can only give rise to single particle-hole excitations which do not contribute to the plasmon lifetime for $v_{\rm F} q\ll \omega \ll 2\varepsilon_{\rm F}$. This in turn implies that $\Im m[\chi_{jj_n}(q,\omega)] = 0$ in the regime of interest (here ${\hat {\bm j}}_{n,{\bm q}}$ is the $n$-th order contribution to ${\hat {\bm j}}'_{\bm q}$).  However, ${\hat {\bm j}}_{1, {\bm q}}$ is a {\it two-particle} operator, implying that  $\Im m[\chi_{j_1j_1}(q,\omega)] \neq 0$ in the regime of interest.
After a lengthy calculation~(see appendix) we arrive at the following expression for the first-order correction to the longitudinal current operator ({\it i.e.}, the current projected along the ${\hat {\bm q}}$ direction):
\begin{equation}
{\hat {\bm q}}\cdot{\hat{\bm  j}}_{1, {\bm q}} = \frac{1}{2}\sum_{{\bm q}'}v_{{\bm q}'}
\left[{\hat \Upsilon}_{{\bm q}, {\bm q}'} {\hat n}_{-{\bm q}'} + {\hat n}_{{\bm q}'} {\hat \Upsilon}_{{\bm q}, - {\bm q}'} \right]~,
\end{equation}
where $v_{{\bm q}'}=2\pi e^2/(\epsilon q')$ is the 2D Fourier transform of the Coulomb interaction and
\begin{eqnarray} \label{eq:Upsilon_definition}
{\hat \Upsilon}_{{\bm q}, {\bm q}'} &=& \sum_{\alpha} \left\{
\frac{v_{\rm F} q}{\omega^2} \left[\frac{q_y'^2}{q'^2} \frac{q'_\alpha}{k_{\rm F}}
-2 \frac{q'_x}{k_{\rm F}} \left(1 - \frac{q'^2}{4 k_{\rm F}^2} \right)\delta_{\alpha, x} 
\right]
\right.
\nonumber\\
&+&
\left.
\frac{q'^2}{4 v_{\rm F} k_{\rm F}^3} \delta_{\alpha, x} \right\} {\hat j}_{{\bm q}',\alpha} \equiv \sum_{\alpha}\Gamma_\alpha({\bm q}, {\bm q}'){\hat j}_{{\bm q}',\alpha}~.
\end{eqnarray}
Here the index $\alpha$ runs over the Cartesian components $x$ (parallel to ${\hat {\bm q}}$) and $y$ (perpendicular to ${\hat {\bm q}}$). The main differences between Eq.~(\ref{eq:Upsilon_definition}) and the corresponding expression for an ordinary 2D electron gas (EG) are (i) the factor $1-q'^2/(4 k_{\rm F}^2)$, which suppresses backscattering at the Fermi surface, and (ii) the last term in curly brackets which remains finite even in the $q\to 0$ limit. Both of them are peculiar to graphene and are intimately related to the chirality of the low-energy MDF model~\cite{castroneto_rmp_2009,kotov_rmp_2012}. See also in appendix for more details.
The next simplification is suggested by the analysis of the Feynman graphs contributing to the noninteracting spectrum of ${\hat {\bm j}}'_{\bm q}$.  These are shown in Fig.~\ref{fig:one}. Because ${\hat {\bm j}}'_{\bm q}$ is (in our approximation) a two-particle operator, these diagrams have {\it four} vertices, one for each creation-annihilation pair.  We see that the disconnected graphs contain two independent sums over the $N_{\rm f}$ electron flavors whereas the connected ones contain only one such sum.  We conclude that the disconnected graphs dominate in the large-$N_{\rm f}$ limit (see the discussion in appendix). The final formula for the spectrum of ${\hat {\bm j}}'_{\bm q}$, which is exact to second order in e-e interactions and in the large-$N_{\rm f}$ limit, has the intuitively appealing form of a convolution of two single-particle spectra:
\begin{widetext}
\begin{eqnarray} \label{eq:chi_rhorho_mode_decoupling_def}
\Im m[\chi_{nn}(q,\omega)]&=&
- \frac{q^2}{\omega^2} \sum_{\alpha,\beta}\int \frac{d^2{\bm q}'}{(2\pi)^2} v_{{\bm q}'}^2
\int_0^\omega \frac{d\omega'}{\pi} \Big\{ \Gamma_\alpha({\bm q},{\bm q}') \Gamma_{\beta}(-{\bm q},-{\bm q}')
\Im m[\chi^{(0)}_{nn}(q',\omega')]\Im m[\chi^{(0)}_{j_\alpha j_\beta}({\bm q}',\omega-\omega')]
\nonumber\\
&+&
\Gamma_\alpha({\bm q},{\bm q}') \Gamma_\beta(-{\bm q},{\bm q}')
\Im m[\chi^{(0)}_{n j_\alpha}(-{\bm q}',\omega')]~\Im m[\chi^{(0)}_{n j_\beta}({\bm q}',\omega-\omega')]\Big\}~.
\end{eqnarray}
\end{widetext}
In Eq.~(\ref{eq:chi_rhorho_mode_decoupling_def}), $\chi^{(0)}_{nn}(q,\omega)$, $\chi^{(0)}_{j_\alpha j_\beta}({\bm q},\omega)$, and $\chi^{(0)}_{n j_\alpha}({\bm q},\omega)$ are the {\it non-interacting} density-density, current-current, and density-current response functions of a 2D gas of MDFs. The integrals in Eq.~(\ref{eq:chi_rhorho_mode_decoupling_def}) can be carried out analytically with the help of known formulas for these response functions~\cite{DiracplasmonsRPA}. The quantities $\{\Gamma_\alpha({\bm q},{\bm q}'), \alpha = x,y\}$ have been introduced in Eq.~(\ref{eq:Upsilon_definition}). The plasmon lifetime is then derived from Eq.~(\ref{eq:plasmonlifetime}).

The final result can be cast (after restoring $\hbar$) into the following elegant form:
\begin{equation}\label{eq:final-lifetime-second-order}
\Gamma_{\rm p}(q) = \frac{\varepsilon_{\rm F}}{\hbar}{\cal A}_{N_{\rm f}}(\alpha_{\rm ee}) \left(\frac{q}{k_{\rm F}}\right)^2~,
\end{equation}
where ${\cal A}_{N_{\rm f}}(\alpha_{\rm ee}) = N_{\rm f}\alpha_{\rm ee}^2 f(N_{\rm f}\alpha_{\rm ee})$ and $f(x)= [15 x^3 - 15 x^2  - 52 x + 42 - 3(5 x^4 - 24 x^2 + 16){\rm arccoth}(1 + x)]/(288 \pi)$. In Eq.~(\ref{eq:final-lifetime-second-order}) we have introduced the dimensionless parameter $\alpha_{\rm ee} = e^2/(\hbar v_{\rm F} \epsilon)$, which measures the strength of e-e interactions relative to the kinetic energy when the low-energy MDF limit is taken~\cite{castroneto_rmp_2009,kotov_rmp_2012}. Here $v_{\rm F} \sim 10^6~{\rm m}/{\rm s}$ is the Fermi velocity. For a flake on a typical substrate like ${\rm SiO}_2$~\cite{novoselov_naturemater_2007,castroneto_rmp_2009,kotov_rmp_2012} or h-${\rm BN}$~\cite{dean_naturenano_2010}, $\alpha_{\rm ee} < 1$, therefore justifying a perturbative treatment of ${\hat {\cal H}}_{\rm ee}$. Only suspended samples~\cite{bolotin_ssc_2008} ($\alpha_{\rm ee} \sim 2.2$) are formally outside the perturbative regime. The dependence of ${\cal A}_{N_{\rm f}}(\alpha_{\rm ee})$ on $\alpha_{\rm ee}$ beyond $\alpha^2_{\rm ee}$, which is encoded into the function $f(x)$ evaluated at $x = N_{\rm f}\alpha_{\rm ee}$, stems from the use of a statically-screened e-e interaction~\cite{abedinpour_prb_2011}, which is needed to cure infrared divergences associated with the Coulomb interaction.

\section{Results and discussion}
\begin{figure}[t]
\begin{center}
\tabcolsep=0cm
\begin{tabular}{c}
\includegraphics[width=1.0\linewidth]{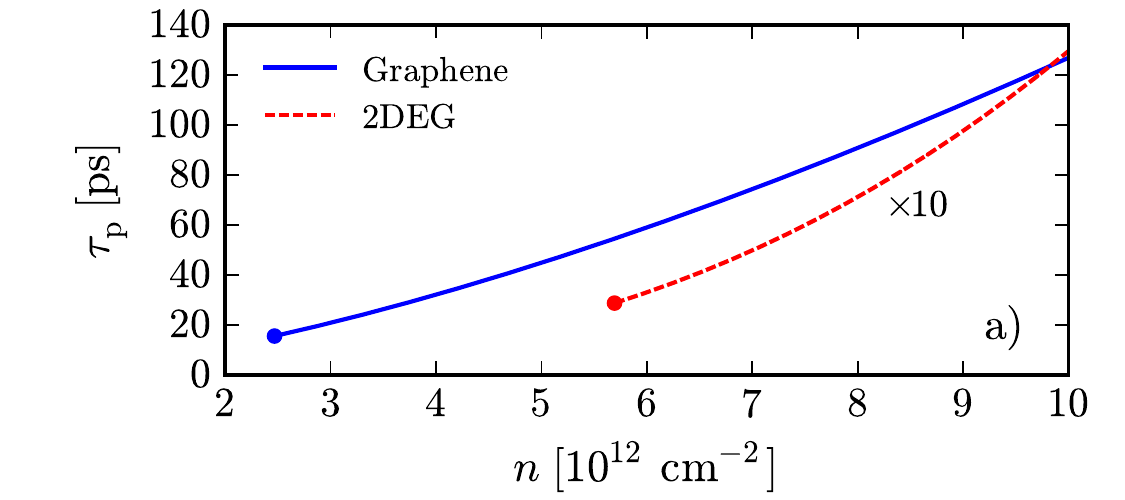}\\
\includegraphics[width=1.0\linewidth]{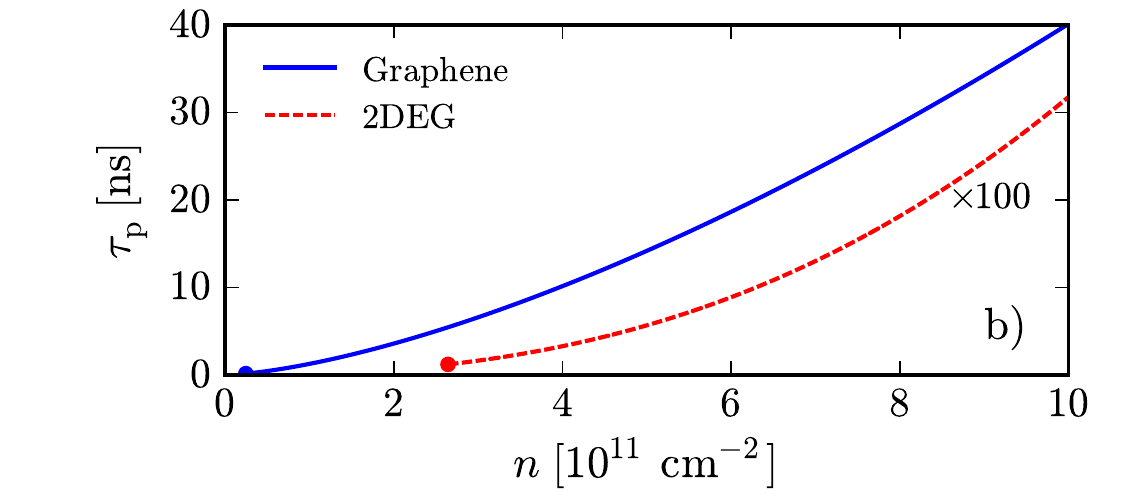}
\end{tabular}
\caption{(Color online) The intrinsic Dirac plasmon lifetime $\tau_{\rm p}(q_1)$ is plotted as a function of electron density $n$ and for a fixed photon energy $\hbar\omega_{\rm ph}$. The (blue) solid line refers to $\alpha_{\rm ee} = 0.9$. The (red) dashed line refers to a 2DEG in a GaAs quantum well. The  intrinsic lifetime of a 2DEG plasmon is much shorter: the dashed curves have been multiplied by large enhancement factors to fit into the frames of the figures. Different panels refer to different values of the photon energy: in panel a) we have set $\hbar \omega_{\rm ph} = 112~{\rm meV}$ corresponding to mid-infrared plasmons; in panel b) $\hbar\omega_{\rm ph} = 11.2~{\rm meV}$ corresponding to Terahertz plasmons. Note the difference in the scales of horizontal and vertical axes between the two panels.
\label{fig:two}}
\end{center}
\end{figure}

In Fig.~\ref{fig:two} we plot the DP lifetime $\tau_{\rm p}(q)$ as calculated from Eq.~(\ref{eq:final-lifetime-second-order}). Following Ref.~[\onlinecite{fei_nature_2012}], this quantity has been plotted for $q$ equal to the plasmon wave number, $q_1/k_{\rm F} = (2\alpha_{\rm ee})^{-1}(\hbar \omega_{\rm ph}/\varepsilon_{\rm F})^2$ for a fixed photon energy $\hbar \omega_{\rm ph}$. As density decreases $q_1/k_{\rm F}$ increases: filled circles in Fig.~\ref{fig:two} refer to the value of doping such that $(q_1/k_{\rm F})_{\rm max} =0.2$. From this figure we clearly see that the intrinsic DP lifetime can be of the order of $20$-$120~{\rm ps}$ for mid-infrared plasmons and of tens of ns for Terahertz plasmons. For the sake of comparison, in Fig.~\ref{fig:two} we have also plotted  the intrinsic lifetime of a plasmon in an ordinary 2DEG hosted in a GaAs quantum well. In this case filled circles refer to the value of doping such that $(q_1/k_{\rm F})_{\rm max} =0.4$. Clearly, DPs have a much longer intrinsic lifetime. This difference stems from the chirality factor which characterizes the electron wave functions in a graphene sheet. As well known~\cite{novoselov_naturemater_2007,castroneto_rmp_2009,kotov_rmp_2012}, this factor suppresses backscattering at the Fermi surface therefore enhancing the DP intrinsic lifetime with respect to that of a plasmon in an ordinary 2DEG.

\begin{figure}[t]
\begin{center}
\tabcolsep=0cm
\begin{tabular}{c}
\includegraphics[width=1.0\linewidth]{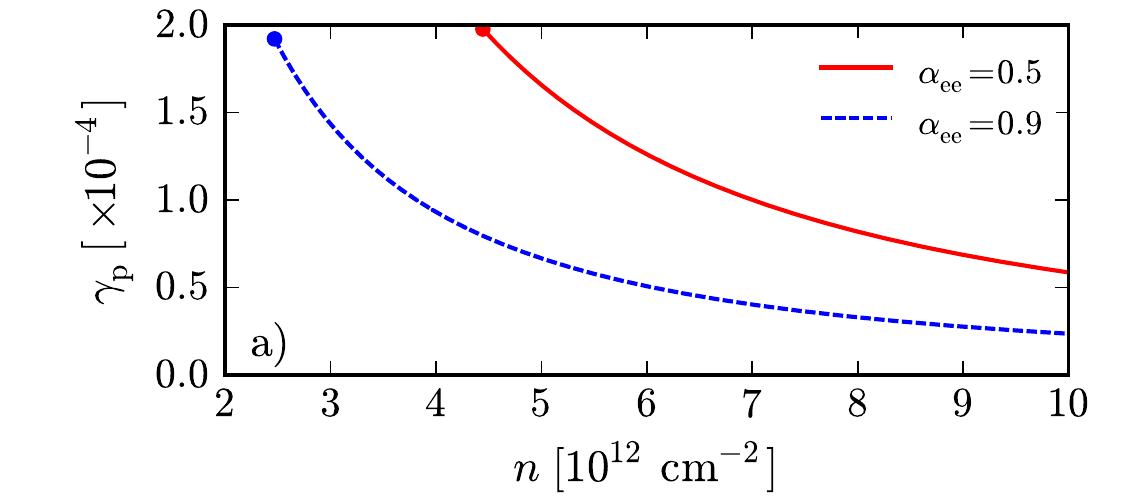}\\
\includegraphics[width=1.0\linewidth]{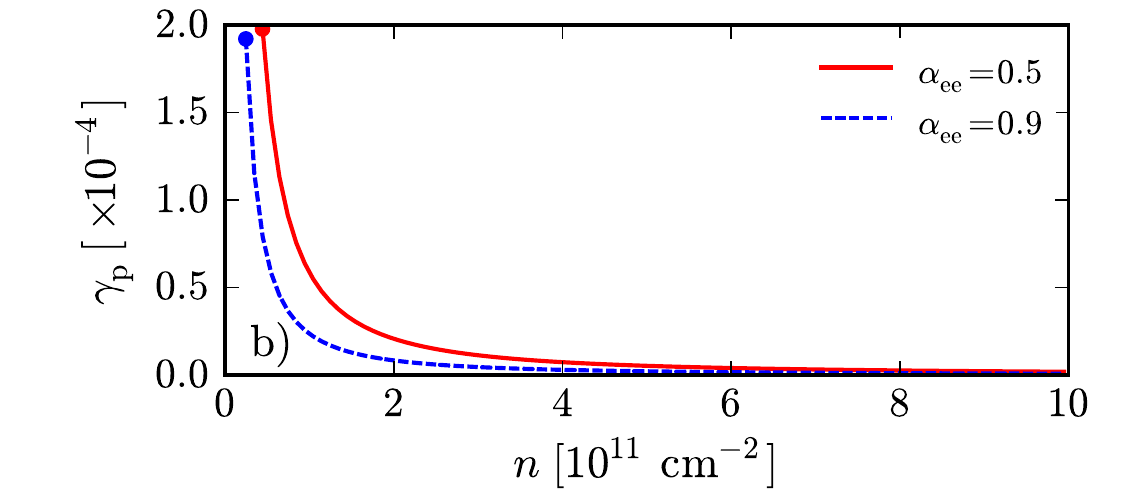}
\end{tabular}
\caption{(Color online) The intrinsic Dirac plasmon damping rate $\gamma_{\rm p}(q_1)$ is plotted as a function of electron density $n$ and for a fixed photon energy $\hbar\omega_{\rm ph}$. In this figure different curves refer to different values of the graphene fine-structure constant $\alpha_{\rm ee}$. As in Fig.~\ref{fig:two}, panel a) [panel b)] refers to mid-infrared [Terahertz] plasmons.\label{fig:three}}
\end{center}
\end{figure}

In Fig.~\ref{fig:three} we plot the DP intrinsic inverse quality factor,
\begin{equation}\label{eq:dampingrate}
\gamma_{\rm p}(q) = \frac{\Gamma_{\rm p}(q)}{\omega_{\rm p}(q)} = \sqrt{2}\frac{{\cal A}_{N_{\rm f}}(\alpha_{\rm ee})}{\sqrt{N_{\rm f}\alpha_{\rm ee}}}\left(\frac{q}{k_{\rm F}}\right)^{3/2}~,
\end{equation}
calculated at $q = q_1$ and as a function of doping.  Notice that our $\gamma_{\rm p}$ is one half of the $\gamma_{\rm p}$ defined in Ref.~[\onlinecite{fei_nature_2012}].  From Eq.~(\ref{eq:dampingrate}) we clearly see that 
$\gamma_{\rm p}(q_1) \propto (\hbar\omega_{\rm ph}/\varepsilon_{\rm F})^3$.
From panel a) we note that, in the range of densities explored in Ref.~[\onlinecite{fei_nature_2012}], the dependence of $\gamma_{\rm p}$ on doping is weak. Although this is in agreement with Ref.~[\onlinecite{fei_nature_2012}], the numerical value we find for the DP inverse quality factor in the mid infrared ($\gamma_{\rm p} \approx 10^{-4}$) is much smaller than the measured value ($\gamma_{\rm p} \approx 10^{-1}$). We are therefore led to conclude that the experiments in Refs.~[\onlinecite{fei_nature_2012,chen_nature_2012}] are far from the intrinsic regime where many-body effects would be dominant. (For a discussion of finite-temperature effects, see App.~\ref{sect:SM_finiteT}.)

Lastly, let us compare our findings for the DP damping rate, Eq.~(\ref{eq:dampingrate}), with the background of optical absorption, $\Re e[\sigma_0(\omega)]$ where $\sigma_0(\omega)\equiv \lim_{q\to0}\sigma(q,\omega)$ is the optical conductivity, calculated at $q=0$ and for frequencies  in the single-particle gap~\cite{universal} $\hbar\omega < 2\varepsilon_{\rm F}$.  Making use of the relation~\cite{Pines_and_Nozieres,Giuliani_and_Vignale}  $\sigma(q,\omega) =ie^2\omega\chi_{nn}(q,\omega)/q^2$ it is easy to show that
\be
\gamma_{\rm p}(q)=\left.\frac{\Re e[\sigma(q,\omega)]}{2 \Im m[\sigma(q,\omega)]}\right\vert_{\omega = \omega_{\rm p}(q)}~,
\ee
suggesting that in the $q \to 0$ limit $\gamma_{\rm p}$ is  linked to the ratio of the real part to the imaginary part of the optical conductivity.  However, this suggestion turns out to be incorrect, because the small-$q$ behavior of $\Re e[\sigma(q,\omega_{\rm p}(q))]$ is different from the  small-$q$ behavior of  $\Re e[\sigma_0(\omega_{\rm p}(q))]$. To second order in $\alpha_{\rm ee}$ a careful calculation shows that~\cite{footnotebeyondquadratic}
\begin{equation}\label{eq:final-absorption-second-order}
\Re e[\sigma_0(\omega \ll 2\varepsilon_{\rm F}/\hbar)] \equiv \sigma_1(\omega) = \frac{2 \hbar^3 {\cal D}_0 {\cal B}_{N_{\rm f}}(\alpha_{\rm ee})}{\pi \varepsilon^3_{\rm F}}\omega^2~,
\end{equation}
where ${\cal B}_{N_{\rm f}}(\alpha_{\rm ee}) = N_{\rm f}\alpha_{\rm ee}^2 g(N_{\rm f}\alpha_{\rm ee})$ and 
$g(x) = [3 x^2 + 3 x -2 - 3 x^2 (2 + x) {\rm arccoth}(1 + x)]/[96 \pi (2 + x)]$. 
In the same range of energies, $\Im m[\sigma_0(\omega)] \equiv \sigma_2(\omega) = {\cal D}_0/\pi \omega$. To second order in $\alpha_{\rm ee}$ we therefore find
\begin{equation}\label{eq:dimensionlessabsorption}
\left.\frac{\sigma_1(\omega)}{\sigma_2(\omega)}\right\vert_{\omega = \omega_{\rm ph}} = 2 {\cal B}_{N_{\rm f}}(\alpha_{\rm ee}) \left(\frac{\hbar\omega_{\rm ph}}{\varepsilon_{\rm F}}\right)^3~.
\end{equation}
Note that Eq.~(\ref{eq:dimensionlessabsorption}) has the same dependence on photon energy and density as Eq.~(\ref{eq:dampingrate}), when the latter is evaluated at $q=q_1$. The functional dependence of $\sigma_1/\sigma_2$ on $\alpha_{\rm ee}$ is different, though, and, in particular, $\sigma_1/\sigma_2$ can be smaller than, comparable to, or larger than $\gamma_{\rm p}$ depending on the value of $\alpha_{\rm ee}$---see Fig.~\ref{fig:SM_three} in appendix.

In summary, we have calculated the intrinsic Dirac plasmon lifetime as solely due to electron-electron interactions---Eq.~(\ref{eq:final-lifetime-second-order})---and the background of optical absorption below the single-particle threshold---Eq.~(\ref{eq:final-absorption-second-order}). Suppressed backscattering due to the chiral nature of the eigenstates of the massless Dirac fermion Hamiltonian yields plasmon lifetimes in graphene which are much longer than the corresponding counterparts in ordinary 2D electron gases. Our calculations demonstrate that current samples~\cite{fei_nature_2012,chen_nature_2012} are not yet in the intrinsic regime. Graphene sheets on h-${\rm BN}$~\cite{dean_naturenano_2010} or suspended samples~\cite{bolotin_ssc_2008} offer the opportunity to reach the intrinsic regime, where our theoretical predictions can be tested.

\section{Acknowledgements}
A.P. and G.V. were supported by the BES Grant DE-FG02-05ER46203. 
M.C. and M.P. acknowledge support by MIUR through the program ``FIRB - Futuro in Ricerca 2010" - Project PLASMOGRAPH (Grant No. RBFR10M5BT).

\appendix

\section{The model}
\label{sect:SM_model}
The graphene tight-binding (TB) Hamiltonian with nearest-neighbor hopping~\cite{castroneto_rmp_2009,Bena_NJP_2009} (from now on $\hbar = 1$) is
\begin{eqnarray} \label{eq:SM_non_int_H}
{\hat {\cal H}}_0 = \sum_{{\bm k}, \alpha, \beta} {\hat \psi}^\dagger_{{\bm k},\alpha} ({\bm f}_{{\bm k}} \cdot {\bm \sigma}_{\alpha\beta}) {\hat \psi}_{{\bm k},\beta}
~,
\end{eqnarray}
where the operator ${\hat \psi}^{(\dagger)}_{{\bm k},\alpha}$ annihilates (creates) an electron with momentum ${\bm k}$ and sublattice index $\alpha=A, B$. 
The vector ${\bm f}_{\bm k}$ is defined as~\cite{Bena_NJP_2009}
\begin{eqnarray} \label{eq:SM_f_vec}
{\bm f}_{{\bm k}} = -t \sum_{i=1}^3 \left(\Re e\left[e^{-i {\bm k}\cdot{\bm \delta}_i}\right], -\Im m\left[e^{-i {\bm k}\cdot{\bm \delta}_i}\right]\right)
~.
\end{eqnarray}
Here $t\simeq 2.8~{\rm eV}$ is the nearest-neighbor tunneling amplitude and ${\bm \delta}_i$ are the vectors which connect an atom (belonging to the $A$ sublattice, say) to all of its three nearest neighbors 
atoms (belonging to the $B$ sublattice): ${\bm \delta}_1 = a \sqrt{3} {\hat {\bm x}}/2 + a {\hat {\bm y}}/2$, ${\bm \delta}_2 = -a \sqrt{3} {\hat {\bm x}}/2 + a{\hat {\bm y}}/2$ and ${\bm \delta}_3 = - a{\hat {\bm y}}$ (where $a \simeq 1.42$~\AA~ is the Carbon-Carbon distance). The eigenvalues of the Hamiltonian~(\ref{eq:SM_non_int_H}) are $\varepsilon_{{\bm k},\lambda} = \lambda |{\bm f}_{\bm k}|$, with $\lambda=\pm$. The momentum sum in Eq.~(\ref{eq:SM_non_int_H}) is restricted to the first Brillouin zone (BZ). All momentum sums in what follows will share this restriction. Finally, the Pauli matrices $\sigma^i_{\alpha\beta}$ ($i=x,y,z$) distinguish the two sites of the unit cell. The non-interacting continuum massless Dirac fermion (MDF) model is obtained from Eq.~(\ref{eq:SM_non_int_H}) in the limit $a\to 0$, keeping the product $t a$ constant. Note that in this limit ${\bm f}_{{\bm K}+{\bm k}} \to v_{\rm F} {\bm k}$ with $v_{\rm F} = 3 t a/2 \sim 10^6 ~{\rm m/s}$ the Fermi velocity.

Defining ${\hat c}_{{\bm k},\lambda}$ and ${\hat c}^\dagger_{{\bm k},\lambda}$ as operators in the eigenstate representation, 
Eq.~(\ref{eq:SM_non_int_H}) can be rewritten as 
$
{\hat {\cal H}}_0 = \sum_{{\bm k}, \lambda} \varepsilon_{{\bm k},\lambda} {\hat c}^\dagger_{{\bm k},\lambda} {\hat c}_{{\bm k},\lambda}
$.
In the same representation the Hamiltonian which describes Coulomb interactions between density fluctuations reads as follows:
\begin{eqnarray} \label{eq:SM_interaction_H}
{\hat {\cal H}}_{\rm ee} = \frac{1}{2} \sum_{{\bm q}} v_{\bm q} {\hat n}_{\bm q} {\hat n}_{-{\bm q}}
~,
\end{eqnarray}
where the density operator is~\cite{Bena_NJP_2009}
\begin{eqnarray} \label{eq:SM_density_op}
{\hat n}_{\bm q} &=& \sum_{{\bm k},\lambda,\lambda'} {\cal D}_{\lambda\lambda'}({\bm k}-{\bm q}/2, {\bm k}+{\bm q}/2) {\hat c}^\dagger_{{\bm k}-{\bm q}/2,\lambda} {\hat c}_{{\bm k}+{\bm q}/2,\lambda'}
~.
\nonumber\\
\end{eqnarray}
In writing Eq.~(\ref{eq:SM_interaction_H}) we have neglected a one-body operator proportional to the total number of particles, which is necessary to avoid self-interactions~\cite{Giuliani_and_Vignale}. 

We are interested in the lifetime of the plasmon mode outside the particle-hole continuum. This quantity is determined by two-particle excitations only, which are generated by {\it two-body} operators. 
Note also that in Eq.~(\ref{eq:SM_interaction_H}) $v_{\bm q}$ is the {\it discrete} Fourier transform of the real-space Coulomb interaction, which is a periodic function of the reciprocal-lattice vectors. Finally, in Eq.~(\ref{eq:SM_density_op}) we defined the ``density vertex"
\begin{eqnarray} \label{eq:SM_D_element}
{\cal D}_{\lambda\lambda'} ({\bm k}, {\bm k}') =
\frac{e^{i(\theta_{\bm k}-\theta_{{\bm k}'})/2} + \lambda\lambda' e^{-i(\theta_{\bm k}-\theta_{{\bm k}'})/2}}{2}
\end{eqnarray}
with $\theta_{\bm k} = {\rm Arg}[f_{{\bm k},{\rm x}}+ i f_{{\bm k},{\rm y}}]$. Here $\{f_{{\bm k}, i}, i=x,y\}$ denote the Cartesian components of the vector ${\bm f}_{\bm k}$.
Note that in the continuum limit and for small ${\bm k}$, $\theta_{{\bm K}+{\bm k}} \to \varphi_{\bm k}$. Here $\varphi_{\bm k}$ is the angle between ${\bm k}$ and the ${\hat {\bm x}}$-axis.

For future purposes we also define the ``pseudospin-density" vertices
\begin{eqnarray} \label{eq:SM_S_x_element}
{\cal S}^{(x)}_{\lambda\lambda'} ({\bm k}, {\bm k}') &=& \frac{\lambda' e^{i(\theta_{\bm k}+\theta_{{\bm k}'})/2} + \lambda e^{-i(\theta_{\bm k}+\theta_{{\bm k}'})/2}}{2}
~,
\end{eqnarray}
and
\begin{eqnarray} \label{eq:SM_S_y_element}
{\cal S}^{(y)}_{\lambda\lambda'} ({\bm k}, {\bm k}') &=& \frac{\lambda' e^{i(\theta_{\bm k}+\theta_{{\bm k}'})/2} - \lambda e^{-i(\theta_{\bm k}+\theta_{{\bm k}'})/2}}{2i}
~.
\end{eqnarray}

In what follows we will concentrate on a doped graphene sheet (``Fermi liquid" regime). For the sake of definiteness we assume that the Fermi energy satisfies $\varepsilon_{\rm F} > 0$. 
Results for $\varepsilon_{\rm F} < 0$ can be recovered by appealing to the particle-hole symmetry of the model defined by Eqs.~(\ref{eq:SM_non_int_H}) and~(\ref{eq:SM_interaction_H}).

\section{The continuity equation}
We now show that, within the TB model defined by Eqs.~(\ref{eq:SM_non_int_H}) and~(\ref{eq:SM_interaction_H}), the density-density and longitudinal current-current response functions satisfy the usual continuity equation~\cite{Giuliani_and_Vignale}. 

Let us first note that, since the momentum sum on the right-hand side of Eq.~(\ref{eq:SM_density_op}) is restricted to the first BZ, $[{\hat n}_{\bm q}, {\hat n}_{{\bm q}'}] = 0$. 
The density operator thus commutes with Eq.~(\ref{eq:SM_interaction_H}). We thus get
\begin{eqnarray} \label{eq:SM_continuity_eq}
i \partial_t {\hat n}_{\bm q}
\!\! &=& \!\!
\sum_{{\bm k},\lambda,\lambda'} \sum_j {\hat c}^\dagger_{{\bm k}-{\bm q}/2,\lambda}
{\hat c}_{{\bm k}+{\bm q}/2,\lambda'} 
({\bm f}_{{\bm k}+{\bm q}/2} - {\bm f}_{{\bm k}-{\bm q}/2})_j
\nonumber\\
\!\! &\times& \!\!
{\cal S}^{(j)}_{\lambda\lambda'} ({\bm k}-{\bm q}/2, {\bm k}+{\bm q}/2)
\equiv
-{\bm q}\cdot {\hat {\bm j}}_{\bm q}
~,
\end{eqnarray}
which also defines the longitudinal part of the current operator ${\hat {\bm j}}_{\bm q}$.

Using the fact that the Kubo product
\begin{eqnarray} \label{eq:SM_Kubo}
\langle\langle {\hat A}; {\hat B} \rangle\rangle_\omega = -i \int_0^\infty dt e^{i(\omega+i\eta) t} \langle[{\hat A}(t),{\hat B}]\rangle
~,
\end{eqnarray}
satisfies the following identities
\begin{eqnarray} 
\langle\langle {\hat A}; {\hat B} \rangle\rangle_\omega &=& \frac{1}{\omega} \langle[{\hat A}, {\hat B}]\rangle + \frac{1}{\omega} \langle\langle i\partial_t{\hat A}; {\hat B} \rangle\rangle_\omega
\nonumber\\
&=&
\frac{1}{\omega} \langle[{\hat A}, {\hat B}]\rangle - \frac{1}{\omega} \langle\langle {\hat A}; i\partial_t{\hat B} \rangle\rangle_\omega
~,
\end{eqnarray}
and that the equal-time commutator of two Hermitian operators is an imaginary quantity, we get
\begin{eqnarray} \label{eq:SM_density_density_def}
\Im m \langle\langle {\hat n}_{\bm q}; {\hat n}_{-{\bm q}} \rangle\rangle_\omega = \frac{1}{\omega^2} 
\Im m \langle\langle {\bm q}\cdot {\hat {\bm j}}_{\bm q}; {\bm q}\cdot {\hat {\bm j}}_{-{\bm q}} \rangle\rangle_\omega
~.
\end{eqnarray}
The Kubo product in Eq.~(\ref{eq:SM_Kubo}) is related to the usual linear response function~\cite{proper_chi} $\chi_{\rm AB}(\omega)$ by the relation: $\chi_{\rm AB}(\omega) = \langle\langle {\hat A}; {\hat B} \rangle\rangle_\omega/S$, where 
$S$ is the 2D electron system area. The average $\langle \ldots \rangle$ on the right-hand side of Eq.~(\ref{eq:SM_Kubo}) is taken over the ground state of the system in the presence of electron-electron (e-e) interactions. Since graphene is treated within a two-band model, the response functions in Eq.~(\ref{eq:SM_density_density_def}) are scalars rather than matrices (see Appendix~7 in Ref.~[\onlinecite{Giuliani_and_Vignale}]) and contain contributions from both intra- and inter-band transitions. All the crystalline effects due to the presence of bands other than $\pi$ and $\pi^\star$ are assumed to be negligible.

\section{The canonical transformation} \label{sect:SM_renormalized_H}
In this Section we reduce the evaluation of Eq.~(\ref{eq:SM_density_density_def}) to the calculation of a non-interacting response function whose operators are ``dressed'' by e-e interactions. 

To this end, we introduce the canonical transformation 
\begin{eqnarray} \label{eq:SM_F_tranf}
{\hat {\cal H}}' = e^{i {\hat F}} [{\hat {\cal H}}_0 + {\hat {\cal H}}_{\rm ee}] e^{-i {\hat F}} \equiv {\hat {\cal H}}_0
~.
\end{eqnarray}
Equation~(\ref{eq:SM_F_tranf}) can be solved order by order in perturbation theory. We define ${\hat F} = \openone + {\hat F}_1 + {\hat F}_2 + \ldots$, where ${\hat F}_i$ is the $i$-th order contribution in e-e interactions to ${\hat F}$. The left-hand side of Eq.~(\ref{eq:SM_F_tranf}) becomes
\begin{eqnarray} \label{eq:SM_F_transf_expansion}
{\hat {\cal H}}' &=& {\hat {\cal H}}_0 + {\hat {\cal H}}_{\rm ee} + i [{\hat F}_1, {\hat {\cal H}}_0]
\nonumber\\
&+&
i [{\hat F}_2, {\hat {\cal H}}_0] + i [{\hat F}_1, {\hat {\cal H}}_{\rm ee}] - \frac{1}{2} [{\hat F}_1, [{\hat F}_1, {\hat {\cal H}}_0]]
\nonumber\\
&+&
\ldots
~.
\end{eqnarray}
The transformation outlined in Eq.~(\ref{eq:SM_F_tranf}) is obtained by determining all the ${\hat F}_i$ from the infinite system of operator identities
\begin{eqnarray} \label{eq:SM_F_1_def}
\left\{
\begin{array}{l}
[{\hat {\cal H}}_0, i {\hat F}_1] = {\hat {\cal H}}_{\rm ee}
\vspace{0.1cm}\\
i [{\hat F}_2, {\hat {\cal H}}_0] + i [{\hat F}_1, {\hat {\cal H}}_{\rm ee}] - \frac{1}{2} [{\hat F}_1, [{\hat F}_1, {\hat {\cal H}}_0]] = 0
\vspace{0.1cm}\\
\ldots
\end{array}
\right.
\end{eqnarray}
As it will be clear in what follows, one must determine only ${\hat F}_1$ to compute Eq.~(\ref{eq:SM_density_density_def}) outside the particle-hole continuum to second order in e-e interactions.

The canonical transformation outlined in Eq.~(\ref{eq:SM_F_tranf}) reduces the Kubo product in Eq.~(\ref{eq:SM_Kubo}) to the evaluation of the non-interacting response function $\langle \langle {\hat A}', {\hat B}' \rangle \rangle_{0,\omega}$. The subscript ``0'' means that the average $\langle \ldots \rangle$ in Eq.~(\ref{eq:SM_Kubo}) has to be performed over the ground state on the non-interacting system and that the time evolution is generated by ${\hat {\cal H}}_0$. However, the operators 
${\hat A}' = e^{i {\hat F}} {\hat A} e^{-i {\hat F}}$ and ${\hat B}' = e^{i {\hat F}} {\hat B} e^{-i {\hat F}}$ are now dressed in a complicated fashion by e-e interactions. 

The ``rotated'' current operator can be expanded in powers of the Coulomb interaction as
\begin{eqnarray} \label{eq:SM_j_prime}
{\bm q}\cdot{\hat {\bm j}}_{\bm q}' = {\bm q}\cdot{\hat {\bm j}}_{\bm q} + {\bm q}\cdot{\hat {\bm j}}_{1,{\bm q}} + {\bm q}\cdot{\hat {\bm j}}_{2,{\bm q}} + \ldots
~,
\end{eqnarray}
where ${\bm q}\cdot{\hat {\bm j}}_{1,{\bm q}} = [i {\hat F}_1, {\bm q}\cdot{\hat {\bm j}}_{\bm q}]$, while the right-hand side of Eq.~(\ref{eq:SM_density_density_def}) now becomes
\begin{eqnarray} \label{eq:SM_chi_jj_def}
&& \!\!\!\!\!\!\!\!
\Im m \langle\langle {\bm q}\cdot{\hat {\bm j}}_{\bm q}'; {\bm q}\cdot{\hat {\bm j}}_{-{\bm q}}' \rangle\rangle_{\omega,0} 
\nonumber\\
&=&
-\pi \sum_m 
\langle 0 |{\bm q}\cdot{\hat {\bm j}}_{\bm q}' |m\rangle \langle m |{\bm q}\cdot{\hat {\bm j}}_{-{\bm q}}' |0\rangle  \delta(\omega-\omega_{m0})
~.
\nonumber\\
\end{eqnarray}
Here $|0\rangle$ is the ground state of the non-interacting system, $|m\rangle$ is an excited state and $\omega_{m0}$ is the excitation energy. Equation~(\ref{eq:SM_chi_jj_def}) is valid for zero temperature and for $\omega>0$. Results for $\omega<0$ can be easily obtained by noting that the imaginary part of the linear-response function we are interested in is antisymmetric~\cite{Giuliani_and_Vignale}  under $\omega \leftrightarrow -\omega$.

Apparently both ${\hat {\bm j}}_{1,{\bm q}}$ and ${\hat {\bm j}}_{2,{\bm q}}$ are needed to calculate Eq.~(\ref{eq:SM_chi_jj_def}) to  second order in the Coulomb interaction. However, since the zeroth-order contribution (${\bm q}\cdot{\hat {\bm j}}_{{\bm q}}$) is a one-body operator it can only generate single-pair excitations, whose phase space is limited to the particle-hole continuum. This implies that both the zeroth- and first-order contribution to Eq.~(\ref{eq:SM_chi_jj_def}) are exactly zero. Moreover,
\begin{eqnarray}
\Im m \langle\langle {\bm q}\cdot{\hat {\bm j}}_{\bm q}; {\bm q}\cdot{\hat {\bm j}}_{2,-{\bm q}} \rangle\rangle_{0,\omega}
&=&
\Im m \langle\langle {\bm q}\cdot{\hat {\bm j}}_{2,{\bm q}}; {\bm q}\cdot{\hat {\bm j}}_{-{\bm q}} \rangle\rangle_{0,\omega} 
\nonumber\\
&=&
0
~.
\end{eqnarray}
Thus, to second order in the Coulomb interaction and outside the particle-hole continuum, Eq.~(\ref{eq:SM_chi_jj_def}) becomes
\begin{eqnarray} \label{eq:SM_chi_jj_def_second}
&& \!\!\!\!\!\!\!\!\!
\Im m \langle\langle {\bm q}\cdot{\hat {\bm j}}_{\bm q}'; {\bm q}\cdot{\hat {\bm j}}_{-{\bm q}}' \rangle\rangle_{\omega,0}
\nonumber\\
&=&
-\pi \sum_m 
\langle 0 |{\bm q}\cdot{\hat {\bm j}}_{1,{\bm q}} |m\rangle \langle m |{\bm q}\cdot{\hat {\bm j}}_{1,-{\bm q}} |0\rangle  \delta(\omega+\omega_{m0})
~.
\nonumber\\
\end{eqnarray}
As stated after Eq.~(\ref{eq:SM_j_prime}), only ${\hat F}_1$ is needed to calculate ${\hat {\bm j}}_{1,{\bm q}}$ and to evaluate Eq.~(\ref{eq:SM_chi_jj_def_second}). Since isotropy is restored after taking the low-energy MDF limit, without any lack of generality we can take ${\bm q} = q {\hat {\bm x}}$.

\section{Calculation of \texorpdfstring{${\hat F}_1$}{F1} and \texorpdfstring{${\hat {\bm j}}_{1,{\bm q}}$}{j1}} \label{sect:SM_calculation_F1_j1}
We define (hereafter ${\bm k}_\pm = {\bm k} \pm {\bm q}'/2$)
\begin{eqnarray}
i {\hat F}_1 &\equiv& \frac{1}{2}\sum_{{\bm q}'} v_{{\bm q}'} \sum_{{\bm k},{\bm k}'} \sum_{\lambda,\lambda',\mu,\mu'} {\cal M}_{\lambda,\lambda',\mu,\mu'}({\bm k},{\bm k}',{\bm q}')
\nonumber\\
&\times&
c^\dagger_{{\bm k}_-,\lambda} c_{{\bm k}_+,\lambda'} c^\dagger_{{\bm k}'_+,\mu} c_{{\bm k}'_-,\mu'}
~,
\end{eqnarray}
and we determine ${\cal M}_{\lambda,\lambda',\mu,\mu'}({\bm k},{\bm k}',{\bm q}')$ to satisfy the first of Eqs.~(\ref{eq:SM_F_1_def}). 
The left-hand side of that equality reads
\begin{eqnarray} \label{eq:SM_H_0_F_1_comm}
[{\hat {\cal H}}_0, i {\hat F}_1] &=& \frac{1}{2}\sum_{{\bm q}'} v_{{\bm q}'} \sum_{{\bm k},{\bm k}'} \sum_{\lambda,\lambda',\mu,\mu'} {\cal M}_{\lambda,\lambda',\mu,\mu'}({\bm k},{\bm k}',{\bm q}')
\nonumber\\
&\times&
(\varepsilon_{{\bm k}_-,\lambda} - \varepsilon_{{\bm k}_+,\lambda'} + \varepsilon_{{\bm k}'_+,\mu} - \varepsilon_{{\bm k}'_-,\mu'})
\nonumber\\
&\times&
c^\dagger_{{\bm k}_-,\lambda} c_{{\bm k}_+,\lambda'} c^\dagger_{{\bm k}'_+,\mu} c_{{\bm k}'_-,\mu'}
~.
\end{eqnarray}
Comparing the previous equation with Eq.~(\ref{eq:SM_interaction_H}) we immediately find:
\begin{eqnarray} \label{eq:SM_M_def}
{\cal M}_{\lambda,\lambda',\mu,\mu'}({\bm k},{\bm k}',{\bm q}') = \frac{{\cal D}_{\lambda\lambda'}({\bm k}_-,{\bm k}_+) {\cal D}_{\mu\mu'}({\bm k}'_+,{\bm k}'_-)}{\varepsilon_{{\bm k}_-,\lambda} - \varepsilon_{{\bm k}_+,\lambda'} + \varepsilon_{{\bm k}'_,\mu} - \varepsilon_{{\bm k}'_-,\mu'}}
~.
\nonumber\\
\end{eqnarray}

The operator ${\hat {\bm j}}_{1,{\bm q}}$ is obtained from the definition given after Eq.~(\ref{eq:SM_j_prime}). We remind the reader that our goal is not to compute ${\hat {\bm j}}_{1,{\bm q}}$ {\it per se}, but to calculate Eq.~(\ref{eq:SM_chi_jj_def_second}). In this equation the matrix elements of ${\bm q}\cdot{\hat {\bm j}}_{1,{\bm q}}$ between the states $|0\rangle$ and $|m\rangle$ are multiplied by $\delta(\omega+\omega_{m0})$, which can be used to simplify the expression of ${\hat {\bm j}}_{1,{\bm q}}$. After some straightforward but lengthy algebraic manipulations we obtain
\begin{eqnarray} \label{eq:SM_j_1_element}
{\bm q}\cdot{\hat {\bm j}}_{1,{\bm q}} =  \frac{1}{2}\sum_{{\bm q}'} v_{{\bm q}'}
\left[{\hat \Upsilon}_{{\bm q}, {\bm q}'} {\hat n}_{-{\bm q}'} + {\hat n}_{{\bm q}'} {\hat \Upsilon}_{{\bm q}, - {\bm q}'} \right]
~.
\end{eqnarray}
Here we defined
\begin{eqnarray} \label{eq:SM_Upsilon_def}
{\hat \Upsilon}_{{\bm q},{\bm q}'} = q \sum_{{\bm k},\lambda,\lambda'} {\hat c}^\dagger_{{\bm k}_--{\bm q}/2,\lambda} {\hat c}_{{\bm k}_++{\bm q}/2,\lambda'} {\hat {\bm q}}\cdot{\bm M}_{\lambda,\lambda'} ({\bm k},{\bm q}',{\bm q}) 
~,
\nonumber\\
\end{eqnarray}
where
\begin{eqnarray} \label{eq:SM_MatrEl_def}
&& \!\!\!\!\!\!\!\!\!\!
{\hat {\bm q}}\cdot {\bm M}_{\lambda,\lambda'} ({\bm k},{\bm q}',{\bm q})
\equiv
\nonumber\\
&& \!\!\!\!\!\!\!\!\!\!
\sum_{\rho} \!\!
\left[
\frac{\displaystyle {\cal D}_{\lambda\rho} \left({\bm k}_--\frac{{\bm q}}{2},{\bm k}_+-\frac{{\bm q}}{2}\right) {\cal S}^{(x)}_{\rho\lambda'}\left({\bm k}_+ -\frac{{\bm q}}{2}, {\bm k}_+ +\frac{{\bm q}}{2}\right)}{\omega + \varepsilon_{{\bm k}_++{\bm q}/2,\lambda'} - \varepsilon_{{\bm k}_+-{\bm q}/2,\rho}}
\right.
\nonumber\\
\!\! &-& \!\!
\left.
\frac{\displaystyle {\cal S}^{(x)}_{\lambda\rho}\left({\bm k}_- -\frac{{\bm q}}{2}, {\bm k}_- +\frac{{\bm q}}{2}\right) {\cal D}_{\rho\lambda'} \left({\bm k}_-+\frac{{\bm q}}{2},{\bm k}_++\frac{{\bm q}}{2}\right)}{\omega + \varepsilon_{{\bm k}_- +{\bm q}/2,\rho} - \varepsilon_{{\bm k}_--{\bm q}/2,\lambda}}
\right]
.
\nonumber\\
\end{eqnarray}
To obtain Eq.~(\ref{eq:SM_MatrEl_def}) we approximated
\begin{eqnarray}
{\bm f}_{{\bm k}+{\bm q}/2} - {\bm f}_{{\bm k}-{\bm q}/2} \to v_{\rm F} {\bm q}
~,
\end{eqnarray}
which becomes exact in the continuum limit when ${\bm k}$ is close to the ${\bm K}$ point of the BZ.

\begin{figure}[t]
\begin{center}
\begin{tabular}{c}
\includegraphics[width=0.99\columnwidth]{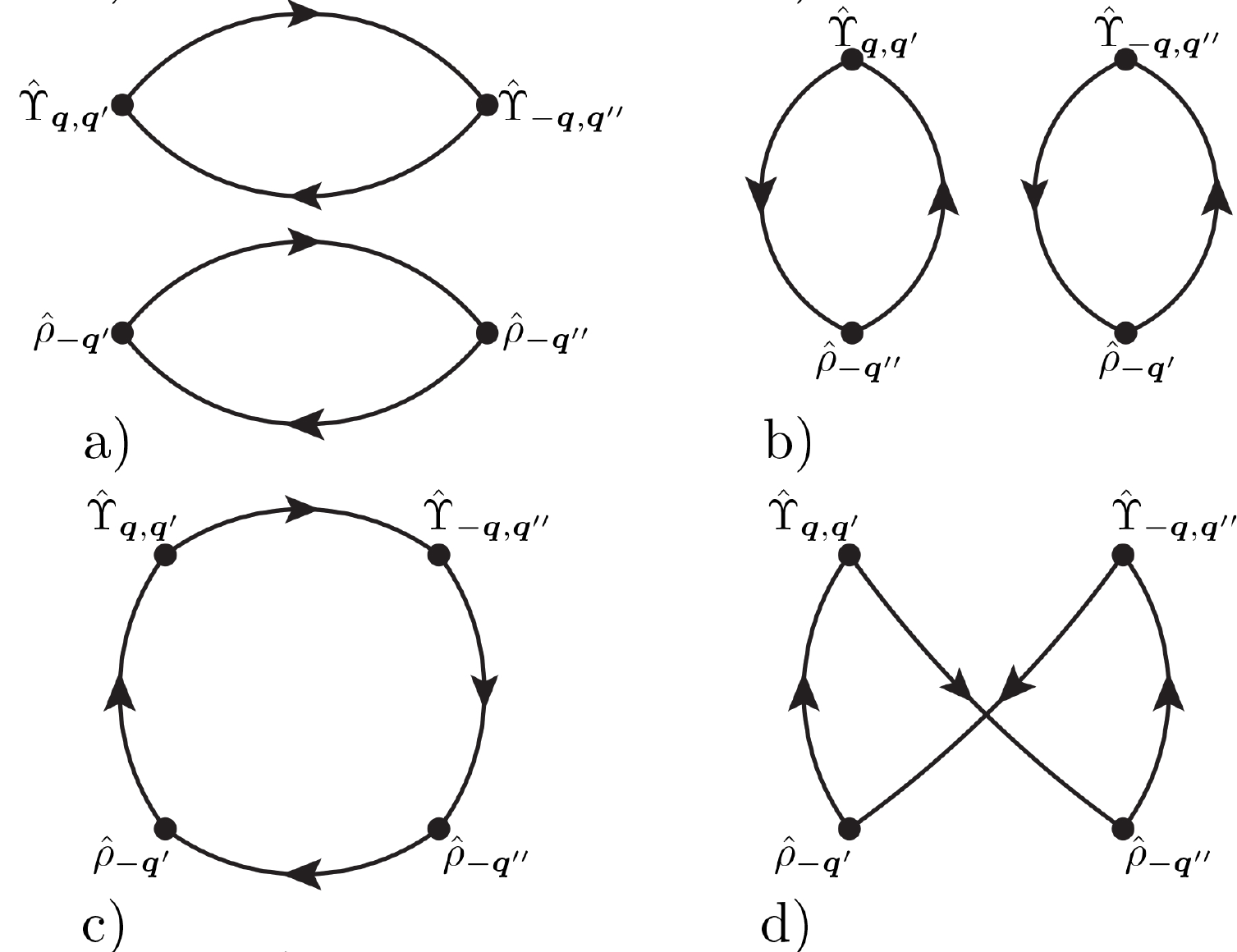}
\end{tabular}
\end{center}
\caption{(Color online) The zeroth order diagrams for the four-body response function. Note that the transformation outlined in Sect.~\ref{sect:SM_renormalized_H} greatly simplifies the problem of evaluating the second-order correction to the linear-response function, reducing it to the calculation of the four diagrams in this figure. Only the diagrams c) and d) contribute to the large-$N_{\rm f}$ expansion of the response function.
}
\label{fig:SM_one}
\end{figure}

After the change of variables ${\bm q}' \to -{\bm q}'$ in the second term on the right-hand side of Eq.~(\ref{eq:SM_j_1_element}), the latter can be rewritten as ${\bm q}\cdot{\hat {\bm j}}_{1,{\bm q}} = \sum_{{\bm q}'} v_{{\bm q}'} {\hat \Upsilon}_{{\bm q}, {\bm q}'} {\hat n}_{-{\bm q}'}$. When this expression is introduced in the Kubo product (\ref{eq:SM_chi_jj_def_second}), the latter admits an expansion in terms of four-point Feynman diagrams, some of which are drawn in Fig.~\ref{fig:SM_one}. In the large-$N_{\rm f}$ limit, one can consider only the sum of the diagrams in Fig.~\ref{fig:SM_one}a) and~b), which reads
\begin{eqnarray} \label{eq:SM_large_N_expansion}
&& \!\!\!\!\!\!\!\!
\Im m \langle\langle {\bm q}\cdot{\hat {\bm j}}_{1,{\bm q}}; {\bm q}\cdot{\hat {\bm j}}_{1,-{\bm q}} \rangle\rangle_{0,\omega}
=
-
\sum_{{\bm q}',{\bm q}''} v_{{\bm q}'} v_{{\bm q}''}
\int_0^\omega \frac{d\omega'}{\pi} 
\nonumber\\
&\times& \!\!\!
\Big[
\Im m \langle\langle {\hat \Upsilon}_{{\bm q},{\bm q'}}; {\hat \Upsilon}_{-{\bm q},{\bm q''}}\rangle\rangle_{0,\omega'}
\Im m \langle\langle {\hat n}_{-{\bm q'}}; {\hat n}_{-{\bm q''}}\rangle\rangle_{0,\omega-\omega'}
\nonumber\\
&+& \!\!\!
\Im m \langle\langle {\hat \Upsilon}_{{\bm q},{\bm q'}}; {\hat n}_{-{\bm q''}}\rangle\rangle_{0,\omega'}
\Im m \langle\langle {\hat n}_{-{\bm q'}}; {\hat \Upsilon}_{-{\bm q},{\bm q''}}\rangle\rangle_{0,\omega-\omega'}
\Big]
~.
\nonumber\\
\end{eqnarray}
Note that this expression coincides with the so-called ``mode-decoupling approximation''~\cite{Giuliani_and_Vignale}. This approximation, although well-known in the electron gas literature, was never demonstrated to be exact in a certain limit before. Finally, we stress that the large-$N_{\rm f}$ expansion in Eq.~(\ref{eq:SM_large_N_expansion}) decouples ${\hat \Upsilon}_{{\bm q},{\bm q'}}$ from the density operator and has the appealing form of a convolution of two single-particle spectra. It is thus possible to study the two operators independently.

\section{Reduction of \texorpdfstring{${\hat \Upsilon}_{{\bm q},{\bm q}'}$}{Upsilon} to a current operator}
\label{sect:SM_Upsilon_manipulation}
In this Section we derive an expression for the operator ${\hat \Upsilon}_{{\bm q},{\bm q}'}$, defined in Eqs.~(\ref{eq:SM_Upsilon_def})-(\ref{eq:SM_MatrEl_def}), which is valid in the limit $v_{\rm F} q \ll \omega \ll \varepsilon_{\rm F}$. In this limit the particle-hole states created by the operator live at the Fermi energy and the band indices on the right-hand side of Eq.~(\ref{eq:SM_MatrEl_def}) are thus constrained to be $\lambda = \lambda' = +$ (recall that $\varepsilon_{\rm F} > 0$).

Expanding the denominators of the two terms on the right-hand side of Eq.~(\ref{eq:SM_MatrEl_def}) we get
\begin{eqnarray} \label{first_term_denominator}
\frac{1}{\omega + \varepsilon_{{\bm k}_\pm+{\bm q}/2,\lambda} - \varepsilon_{{\bm k}_\pm-{\bm q}/2,\rho}} 
&\to&
\delta_{\lambda,\rho} \left[ \frac{1}{\omega} - \frac{q}{\omega^2} \frac{\partial \varepsilon_{{\bm k}_\pm}}{\partial k_x} \right] 
\nonumber\\
&+&
(1-\delta_{\lambda,\rho}) \frac{1}{\omega + 2\lambda\varepsilon_{\rm F}}
\nonumber\\
&+&
{\cal O}(q^2)~.
\end{eqnarray}
Note that we can safely take the limit $\omega \to 0$ in the second term on the right-hand side of Eq.~(\ref{first_term_denominator}). Furthermore, note that the sum in Eq.~(\ref{eq:SM_MatrEl_def}) is carried out on the ``virtual state'' 
$\rho$ which can be either in conduction or valence band, even though the real states (labeled by the band indices $\lambda$ and $\lambda'$) are bound to the Fermi surface. We now define:
\begin{widetext}
\begin{eqnarray} \label{eq:SM_M_intra_def}
{\hat {\bm q}}\cdot{\bm M}_{\rm intra} &\equiv&
\cos\left(\frac{\theta_{{\bm k}_--{\bm q}/2} - \theta_{{\bm k}_+-{\bm q}/2}}{2}\right) \cos(\theta_{{\bm k}_+})
\left[ \frac{1}{\omega} - \frac{v_{\rm F} q}{\omega^2} \cos(\theta_{{\bm k}_+})\right]
\nonumber\\
&-&
\cos\left(\frac{\theta_{{\bm k}_-+{\bm q}/2} - \theta_{{\bm k}_++{\bm q}/2}}{2}\right) \cos(\theta_{{\bm k}_-})
\left[ \frac{1}{\omega} - \frac{v_{\rm F} q}{\omega^2} \cos(\theta_{{\bm k}_-}) \right]
+ {\cal O}(q^2)
~,
\end{eqnarray}
which is obtained from Eq.~(\ref{eq:SM_MatrEl_def}) by taking $\rho=\lambda=\lambda'= +$, and
\begin{eqnarray} \label{eq:SM_M_inter_def}
{\hat {\bm q}}\cdot{\bm M}_{\rm inter} &\equiv&
-\frac{1}{\omega+2\varepsilon_{\rm F}} \sin\left(\frac{\theta_{{\bm k}_--{\bm q}/2} - \theta_{{\bm k}_+-{\bm q}/2}}{2}\right) \sin\left(\frac{\theta_{{\bm k}_+ -{\bm q}/2} + \theta_{{\bm k}_+ +{\bm q}/2}}{2}\right)
\nonumber\\
&-&
\frac{1}{\omega-2\varepsilon_{\rm F}} \sin\left(\frac{\theta_{{\bm k}_-+{\bm q}/2} - \theta_{{\bm k}_++{\bm q}/2}}{2}\right) \sin\left(\frac{\theta_{{\bm k}_- -{\bm q}/2} + \theta_{{\bm k}_- +{\bm q}/2}}{2}\right)
+ {\cal O}(q^2)
~,
\end{eqnarray}
\end{widetext}
which is obtained for $\rho=-$. In Eqs.~(\ref{eq:SM_M_intra_def})-(\ref{eq:SM_M_inter_def}) we have expanded the functions up to linear order in $q$. To ${\cal O}(q^2)$, $M_{\lambda,\lambda'}({\bm k},{\bm q}',{\bm q}) = M_{\rm intra} + M_{\rm inter}$ (the dependence of $M_{\rm intra}$ and $M_{\rm inter}$ on wavevectors and band indices is suppressed for the sake of brevity).

We stress that the subscript ``intra'' [``inter''] in Eq.~(\ref{eq:SM_M_intra_def}) [(\ref{eq:SM_M_inter_def})] refers to the virtual state and {\it not} to the real states, which are constrained to be at the Fermi surface since $\omega \ll \varepsilon_{\rm F}$. The excitations generated by the operator ${\hat \Upsilon}_{{\bm q},{\bm q}'}$ in this limit are indeed always intraband electron-hole pairs. To obtain Eq.~(\ref{eq:SM_M_intra_def}) we used that
\begin{eqnarray}
\cos\left(\frac{\theta_{{\bm k}_\pm -{\bm q}/2} + \theta_{{\bm k}_\pm +{\bm q}/2}}{2}\right) = \cos(\theta_{{\bm k}_\pm}) + {\cal O} (q^2)
~,
\end{eqnarray}
and we approximated
\begin{eqnarray}
\frac{\partial \varepsilon_{{\bm k}_\pm,\lambda}}{\partial k_x} \simeq \lambda v_{\rm F} \cos(\theta_{{\bm k}_\pm})
~.
\end{eqnarray}
The last equality becomes exact in the continuum limit and for ${\bm k}$ close to the ${\bm K}$ point of the BZ.

Let us first consider $M_{\rm intra}$ defined in Eq.~(\ref{eq:SM_M_intra_def}). This expression can be further simplified by noting that
\begin{eqnarray}
\cos\left(\frac{\theta_{{\bm k}_-\pm{\bm q}/2} - \theta_{{\bm k}_+\pm{\bm q}/2}}{2}\right)
\!\! &=& \!\! 
\cos\left(\frac{\theta_{{\bm k}_--{\bm q}/2} - \theta_{{\bm k}_++{\bm q}/2}}{2}\right)
\nonumber\\
\!\! &-& \!\! 
\frac{q}{2} \sin\left(\frac{\theta_{{\bm k}_-} - \theta_{{\bm k}_+}}{2}\right) \frac{\partial \theta_{{\bm k}_\mp}}{\partial k_x}
\nonumber\\
\!\! &+& \!\! 
{\cal O}(q^2)
~,
\end{eqnarray}
which leads to
\begin{eqnarray} \label{eq:SM_M_intra_2}
&& \!\!\!\!\!\!\!\!\!\!
{\hat {\bm q}} \!\cdot\! {\bm M}_{\rm intra}
=
\frac{\cos(\theta_{{\bm k}_+}) - \cos(\theta_{{\bm k}_-})}{\omega}
\cos\!\!\left(\!\frac{\theta_{{\bm k}_--{\bm q}/2} - \theta_{{\bm k}_++{\bm q}/2}}{2}\!\!\right)
\nonumber\\
&+&
\frac{v_{\rm F} q}{\omega^2} [\cos^2(\theta_{{\bm k}_-}) - \cos^2(\theta_{{\bm k}_+})] \cos\left(\frac{\theta_{{\bm k}_-} - \theta_{{\bm k}_+}}{2}\right)
\nonumber\\
&+&
\frac{q}{2\omega} \frac{\partial [\sin(\theta_{{\bm k}_-}) - \sin(\theta_{{\bm k}_-})]}{\partial k_x}
\sin\left(\frac{\theta_{{\bm k}_-} - \theta_{{\bm k}_+}}{2}\right)
\nonumber\\
&+&
{\cal O}(q^2)
~.
\end{eqnarray}
In the first term on the right-hand side of Eq.~(\ref{eq:SM_M_intra_2}) we can approximate
\begin{eqnarray} \label{eq:SM_M_intra_first}
\cos(\theta_{{\bm k}_+}) - \cos(\theta_{{\bm k}_-}) \simeq \frac{q'_x}{k_{\rm F}}
~,
\end{eqnarray}
while the second term on the right-hand side of Eq.~(\ref{eq:SM_M_intra_2}) becomes
\begin{eqnarray} \label{eq:SM_M_intra_second}
&& \!\!\!\!\!\!\!\!
[\cos^2(\theta_{{\bm k}_-}) - \cos^2(\theta_{{\bm k}_+})] \cos\left(\frac{\theta_{{\bm k}_-} - \theta_{{\bm k}_+}}{2}\right)
\nonumber\\
&\simeq&
- 2 \frac{q'_x}{k_{\rm F}} \left(1 - \frac{q'^2}{4 k_{\rm F}^2} \right) \cos\left(\frac{\theta_{{\bm k}_-} + \theta_{{\bm k}_+}}{2}\right) 
~.
\end{eqnarray}
Finally, the derivative in the third term on the right-hand side of Eq.~(\ref{eq:SM_M_intra_2}) is
\begin{eqnarray} \label{eq:SM_M_intra_third}
\frac{\partial [\sin(\theta_{{\bm k}_-}) - \sin(\theta_{{\bm k}_-})]}{\partial k_x}
&\simeq&
- \frac{\partial (q'_y/k_{\rm F})}{\partial k_x}
\nonumber\\
&=&
0
~.
\end{eqnarray}
Equations~(\ref{eq:SM_M_intra_first})-(\ref{eq:SM_M_intra_third}) become exact in the continuum limit, for ${\bm k}$ close to the ${\bm K}$ point of the BZ and for $v_{\rm F} q \ll \omega \ll \varepsilon_{\rm F}$. 

Introducing Eq.~(\ref{eq:SM_M_intra_2}), approximated according to Eqs.~(\ref{eq:SM_M_intra_first})-(\ref{eq:SM_M_intra_third}), back into Eq.~(\ref{eq:SM_Upsilon_def}) we get the ``intraband'' contribution to the operator ${\hat \Upsilon}_{{\bm q},{\bm q}'}$, which reads
\begin{eqnarray} \label{eq:SM_Upsilon_intra_def}
{\hat \Upsilon}^{({\rm intra})}_{{\bm q},{\bm q}'} 
\!\!\! &=& \!\!\!
\left[\frac{v_{\rm F} q'_x}{k_{\rm F}\omega} {\hat n}_{{\bm q}+{\bm q}'}
- 2 \frac{v_{\rm F} q_x}{\omega^2} \frac{q'_x}{k_{\rm F}} \left(1 - \frac{q'^2}{4 k_{\rm F}^2} \right)
j_{{\bm q}',x}
\right]
\nonumber\\
&+&
{\cal O}(q^2)
~.
\end{eqnarray}
Recall that ${\bm q} = q {\hat {\bm x}}$. Here we used that ${\hat {\bm j}}_{\bm q} = v_{\rm F} {\hat {\bm \sigma}}_{\bm q}$ close to the ${\bm K}$ point of the BZ.

Let us now consider $M_{\rm inter}$ defined as in Eq.~(\ref{eq:SM_M_inter_def}). Setting $q=0$ in Eq.~(\ref{eq:SM_M_inter_def}) and then taking the limit $\omega \to 0$
\begin{eqnarray} \label{eq:SM_MatrEl_inter_def}
{\hat{\bm q}} \!\cdot\! {\bm M}_{\rm inter} &=&
\frac{1}{2\varepsilon_{\rm F}}
[\sin(\theta_{{\bm k}_-}) - \sin(\theta_{{\bm k}_+})] \sin\left(\frac{\theta_{{\bm k}_-} - \theta_{{\bm k}_+}}{2}\right)
\nonumber\\
&=&
\frac{1}{\varepsilon_{\rm F}} \sin^2\left(\frac{\theta_{{\bm k}_-} - \theta_{{\bm k}_+}}{2}\right) \cos\left(\frac{\theta_{{\bm k}_-} + \theta_{{\bm k}_+}}{2}\right)
\nonumber\\
&=&
\frac{q'^2}{4 v_{\rm F} k_{\rm F}^3} {\cal S}^{(x)}_{\lambda\lambda'}({\bm k}_-,{\bm k}_+)
~,
\end{eqnarray}
which, as usual, becomes exact in the continuum limit and for ${\bm k}$ close to the ${\bm K}$ point of the BZ. Eq.~(\ref{eq:SM_MatrEl_inter_def}), when introduced into Eq.~(\ref{eq:SM_Upsilon_def}), gives the ``interband'' contribution to the operator ${\hat \Upsilon}_{{\bm q},{\bm q}'}$, {\it i.e.}
\begin{eqnarray} \label{eq:SM_Upsilon_inter_def}
{\hat \Upsilon}^{({\rm inter})}_{{\bm q},{\bm q}'} &=&
\frac{q'^2}{4 v_{\rm F} k_{\rm F}^3} j_{{\bm q}',x}
+{\cal O}(q^2)
~.
\end{eqnarray}
Again, we used the fact that ${\hat {\bm j}}_{\bm q} = v_{\rm F} {\hat {\bm \sigma}}_{\bm q}$ close to the ${\bm K}$ point of the BZ.

Putting Eqs.~(\ref{eq:SM_Upsilon_intra_def}) and~(\ref{eq:SM_Upsilon_inter_def}) together we finally get
\begin{eqnarray} \label{eq:SM_Upsilon_approx}
{\hat \Upsilon}_{{\bm q},{\bm q}'} &=& 
\frac{v_{\rm F} q'_x}{k_{\rm F}\omega} {\hat n}_{{\bm q}+{\bm q}'}
+ {\hat \Upsilon}'_{{\bm q},{\bm q}'}
~.
\end{eqnarray}
Here
\begin{eqnarray}
{\hat \Upsilon}'_{{\bm q},{\bm q}'} = - \left[ 2  \frac{v_{\rm F} q_x}{\omega^2} \frac{q'_x}{k_{\rm F}} \left(1 - \frac{q'^2}{4 k_{\rm F}^2} \right) - \frac{q'^2}{4 v_{\rm F} k_{\rm F}^3} \right] {\hat j}_{{\bm q}',x}
~.
\nonumber\\
\end{eqnarray}
As already stressed, the previous equation becomes exact (i) in the continuum limit taken close to the ${\bm K}$ point of the BZ and (ii) in the limit $v_{\rm F} q \ll \omega \ll \varepsilon_{\rm F}$. 

We can further manipulate the first term on the right-hand side of Eq.~(\ref{eq:SM_Upsilon_approx}). When this is introduced into Eq.~(\ref{eq:SM_j_1_element}) it gives a contribution of the form
\begin{eqnarray} \label{eq:SM_rhorho_manipulation}
&& \!\!\!\!\!\!\!\!
\frac{1}{2\omega k_{\rm F}}\sum_{{\bm q}'} v_{{\bm q}'} 
\left[q'_x{\hat n}_{{\bm q} + {\bm q}'} {\hat n}_{-{\bm q}'} - q'_x {\hat n}_{{\bm q}'} {\hat n}_{{\bm q} - {\bm q}'} \right]
\nonumber\\
&=&
\frac{v_{\rm F} }{2\omega k_{\rm F}} \sum_{{\bm q}'} {\hat n}_{{\bm q} + {\bm q}'} {\hat n}_{-{\bm q}'}
[q'_x v_{{\bm q}'} - (q+q'_x) v_{{\bm q} + {\bm q}'}]
\nonumber\\
&\to&
\frac{v_{\rm F} q}{2\omega k_{\rm F}}\sum_{{\bm q}'} v_{{\bm q}'} \left(\frac{q_x'^2}{q'^2} - 1\right){\hat n}_{{\bm q}'} {\hat n}_{-{\bm q}'}
+ {\cal O}(q^2)
~.
\end{eqnarray}
Here we shifted ${\bm q}' \to {\bm q}+{\bm q}'$ in the term proportional to ${\hat n}_{{\bm q}'} {\hat n}_{{\bm q} - {\bm q}'}$ and we took the small-${\bm q}$ limit in the last line of Eq.~(\ref{eq:SM_rhorho_manipulation}). Finally, using the continuity equation
\begin{eqnarray}
\omega {\hat n}_{{\bm q}'} {\hat n}_{-{\bm q}'} = - {\bm q}'\cdot {\hat {\bm j}}_{{\bm q}'} {\hat n}_{-{\bm q}'} + {\hat n}_{{\bm q}'} {\bm q}'\cdot {\hat {\bm j}}_{-{\bm q}'}
~,
\end{eqnarray}
it is possible to redefine the operator of Eq.~(\ref{eq:SM_Upsilon_approx}) as
\begin{eqnarray} \label{eq:SM_Upsilon_approx_3}
{\hat \Upsilon}_{{\bm q}, {\bm q}'} 
\!\! &=& \!\!
\sum_{\alpha}
\Bigg\{
\frac{v_{\rm F} q_x}{\omega^2} \Bigg[ \frac{q_y'^2}{q'^2} \frac{q'_\alpha}{k_{\rm F}}
-
2 \frac{q'_x}{k_{\rm F}}
\Bigg( 1 - \frac{q'^2}{4 k_{\rm F}^2} \!\Bigg)
\delta_{\alpha,x} 
\Bigg]
\nonumber\\
&+&
\frac{q'^2}{4 v_{\rm F} k_{\rm F}^3} \delta_{\alpha,x} 
\! \Bigg\}
{\hat j}_{{\bm q}',\alpha}
\equiv
\sum_{\alpha} \Gamma_\alpha({\bm q},{\bm q}') {\hat j}_{{\bm q}',\alpha}
~.
\end{eqnarray}
The main differences of Eq.~(\ref{eq:SM_Upsilon_approx_3}) with the analogous equation that can be derived following a similar procedure for an ordinary 2D electron gas are (i) the factor $1-q'^2/(4 k_{\rm F}^2)$ which is due to chirality and suppresses backscattering at the Fermi surface, and (ii) the last term in curly brackets, which is finite even for $q\to 0$. The latter is due to the two-band nature of graphene, which opens the possibility of a {\it virtual} state in valence band even though the real states are at the Fermi energy in conduction band.

Eq.~(\ref{eq:SM_Upsilon_approx_3}) allows us to write Eq.~(\ref{eq:SM_large_N_expansion}) to order $q^2$ as
\begin{widetext}
\begin{eqnarray} \label{eq:SM_chi_rhorho_mode_decoupling_def}
\Im m \langle\langle {\hat n}_{\bm q}; {\hat n}_{-{\bm q}} \rangle\rangle_\omega &=&
- \frac{q^2}{\omega^2} \sum_{\alpha,\beta} \int_{\rm BZ} \frac{d^2{\bm q}'}{(2\pi)^2} v_{{\bm q}'}^2
\int_0^\omega \frac{d\omega'}{\pi} \Big[ \Gamma_\alpha({\bm q},{\bm q}') \Gamma_{\beta}(-{\bm q},-{\bm q}')
\Im m\chi^{(0)}_{nn}(-{\bm q}',\omega') \Im m\chi^{(0)}_{j_\alpha j_{\beta}}({\bm q}',\omega-\omega')
\nonumber\\
&+&
\Gamma_\alpha({\bm q},{\bm q}') \Gamma_{\beta}(-{\bm q},{\bm q}')
\Im m\chi^{(0)}_{n j_\alpha}(-{\bm q}',\omega') \Im m\chi^{(0)}_{n j_{\beta}}({\bm q}',\omega-\omega')
\Big]
~.
\end{eqnarray}
\end{widetext}
Note that the {\it imaginary} parts of the non-interacting current-current [$\Im m\chi^{(0)}_{j_\ell j_{\ell'}}({\bm q},\omega)$], current-density [$\Im m\chi^{(0)}_{j_\ell n}({\bm q},\omega)$] and density-density [$\Im m\chi^{(0)}_{nn}({\bm q},\omega)$] response functions are all cutoff-free in both the tight-binding model and in the continuum limit. Moreover, the ${\bm q}'$-integral is naturally bounded, in the limit of $\omega\to 0$, to $0\leq q' \leq 2k_{\rm F}$. Since no regularization is needed in Eq.~(\ref{eq:SM_chi_rhorho_mode_decoupling_def}) it can be safely evaluated in the continuum limit.

The calculation of the plasmon damping rate $\gamma_{\rm p} (q)$ and of the optical spectrum $\sigma_1(\omega)$ from Eq.~(\ref{eq:SM_chi_rhorho_mode_decoupling_def}), although straightforward, is quite lengthy and will not be reported here. 

\begin{figure}[t!]
\begin{center}
\begin{tabular}{c}
\includegraphics[width=0.99\columnwidth]{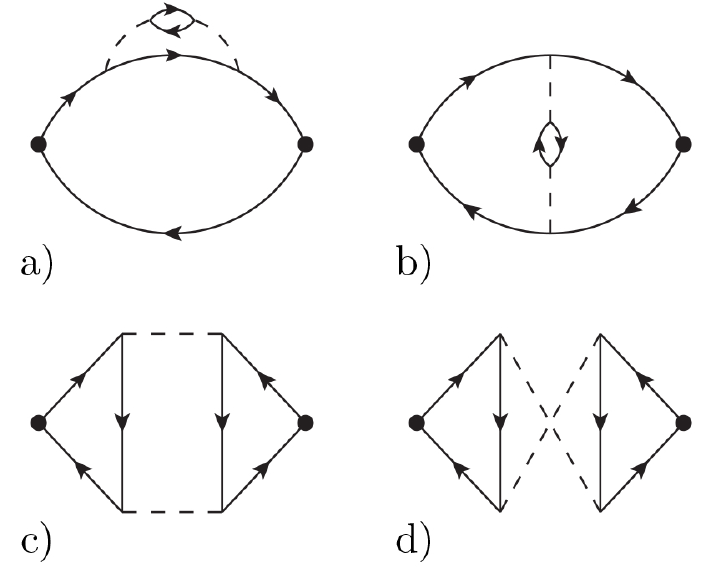}
\end{tabular}
\end{center}
\caption{ (Color online) The large-$N_{\rm f}$ diagrams for the proper density-density response function at second order in the strength of Coulomb interactions.
Panel~a) shows one of the two time-reversal-conjugated self-energy diagrams, while panel b) illustrates the second-order vertex correction. Finally, 
panels~c) and~d) depict two Aslamazov-Larkin-type diagrams. Solid (dashed) lines represent non-interacting Green's functions (e-e interactions). The external vertices (filled dots) are density operators.
}
\label{fig:SM_two}
\end{figure}

In passing, we would like to mention that the ``mode-decoupling" formula (\ref{eq:SM_chi_rhorho_mode_decoupling_def}) yields for the plasmon lifetime the same result that can be calculated from the diagrams for the density-density response function at second order in the strength of e-e interactions and in the large-$N_{\rm f}$ limit. These diagrams are shown in Fig.~\ref{fig:SM_two}.

\section{The plasmon damping rate and the optical spectrum}
%
\begin{figure}[h!]
\begin{center}
\begin{tabular}{c}
\includegraphics[width=0.99\columnwidth]{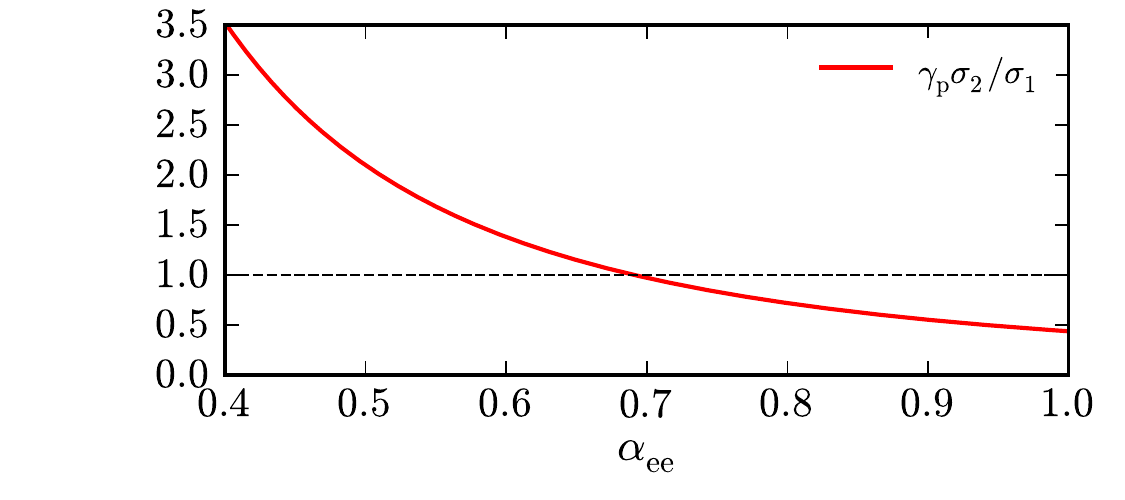}
\end{tabular}
\end{center}
\caption{(Color online) The ratio between $\gamma_{\rm p} (q_1)$ and $\sigma_1(\omega)/\sigma_2(\omega)|_{\omega=\omega_{\rm ph}}$, defined respectively in Eqs.~(\ref{eq:SM_dampingrate}) and~(\ref{eq:SM_dimensionlessabsorption}). The two functions become equal for $\alpha_{\rm ee} \simeq 0.7$.}
\label{fig:SM_three}
\end{figure}

In this Section we comment on the ratio between the plasmon damping rate and the background of optical absorption due to e-e interactions below the single-particle gap. 

Restoring $\hbar$, the damping rate is defined as [see Eq.~(7) in the main text]
\begin{eqnarray}\label{eq:SM_dampingrate}
\gamma_{\rm p}(q_1) = \frac{{\cal A}_{N_{\rm f}}(\alpha_{\rm ee})}{2 \sqrt{N_{\rm f}} \alpha_{\rm ee}^2} \left(\frac{\hbar\omega_{\rm ph}}{\varepsilon_{\rm F}}\right)^3~,
\end{eqnarray}
while the dimensionless absorption spectrum [Eq.~(9) in the main text] as
\begin{equation}\label{eq:SM_dimensionlessabsorption}
\left.\frac{\sigma_1(\omega)}{\sigma_2(\omega)}\right\vert_{\omega = \omega_{\rm ph}} = 2 {\cal B}_{N_{\rm f}}(\alpha_{\rm ee}) \left(\frac{\hbar\omega_{\rm ph}}{\varepsilon_{\rm F}}\right)^3~.
\end{equation}
In Eq.~(\ref{eq:SM_dampingrate}) we have introduced the plasmon wave number $q_1/k_{\rm F} = (2\alpha_{\rm ee})^{-1}(\hbar \omega_{\rm ph}/\varepsilon_{\rm F})^2$. Note that
Eqs.~(\ref{eq:SM_dampingrate}) and~(\ref{eq:SM_dimensionlessabsorption}) share the same dependence on the photon energy and on carrier density. Their functional dependence on the coupling constant $\alpha_{\rm ee}$ is, however, different. In Fig.~\ref{fig:SM_three} we plot the ratio between Eq.~(\ref{eq:SM_dampingrate}) and Eq.~(\ref{eq:SM_dimensionlessabsorption}). We clearly see that these two quantities become equal for $\alpha_{\rm ee} \sim 0.7$.

\section{The plasmon damping rate at finite temperature} \label{sect:SM_finiteT}
%
\begin{figure}[t]
\begin{center}
\begin{tabular}{c}
\includegraphics[width=0.99\columnwidth]{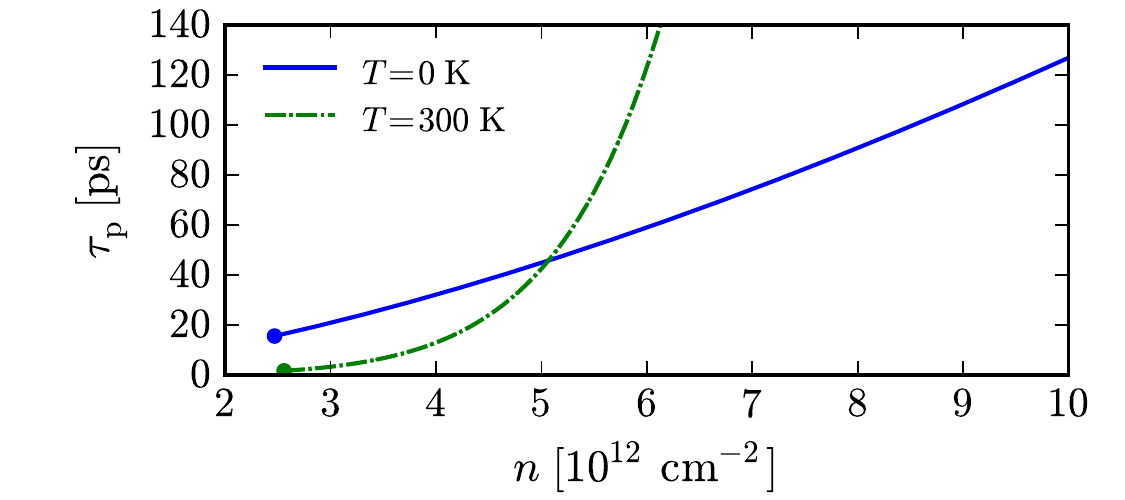}
\end{tabular}
\end{center}
\caption{(Color online) The Dirac plasmon lifetime $\tau_{\rm p}(q_1)$ is plotted as a function of electron density $n$ and for a fixed photon energy $\hbar\omega_{\rm ph} = 112~{\rm meV}$. The (blue) solid line refers to the intrinsic plasmon lifetime calculated at $T=0$ [the same function is plotted in Fig.~2a) of the main text]. The (green) dash-dotted line refers to the RPA plasmon lifetime computed from the finite-temperature Lindhard function~\cite{ramezanali_jphysa_2009} at $T=300~{\rm K}$. Both curves refer to $\alpha_{\rm ee} = 0.9$.}
\label{fig:SM_four}
\end{figure}

All the calculations described in the main text have been performed at zero temperature. Finite-temperature effects introduce additional damping due to the presence of thermally excited quasiparticles. We have estimated this effect using RPA~\cite{ramezanali_jphysa_2009} and the results are plotted in Fig.~\ref{fig:SM_four}. We conclude that the temperature effect is negligible at the typical densities of the experiments of Refs.~[\onlinecite{fei_nature_2012,chen_nature_2012}], but certainly not at lower densities. The thermal broadening of the Dirac plasmon should therefore be considered carefully in any quantitative comparison between theory and experiment.


\begin{thebibliography}{77}
\bibitem{electronholesymmetry}
	We discuss electron doping for the sake of definiteness: $\tau_{\rm p}$ is a particle-hole symmetric quantity.
\end{thebibliography}
\end{document}